\documentclass{JFM-FLM_Au}

\usepackage{siunitx}
\newcommand{\Ek}{\textit{Ek}}
\newcommand{\Ro}{\textit{Ro}}
\renewcommand{\vec}[1]{\boldsymbol{#1}}

\newcommand{\D}{\mathrm{D}}

\lefttitle{S. Peng and A. Bodner}
\righttitle{Perturbed oceanic submesoscales}

\title{A theoretical model for oceanic submesoscales under next-order effects of strain and turbulence}

\author{Shirui Peng\aff{1} \and Abigail Bodner\aff{1,2}}

\affiliation{\aff{1}Department of Earth, Atmospheric and Planetary Sciences, Massachusetts Institute of Technology, Cambridge, MA 02139, USA
\aff{2}Department of Electrical Engineering and Computer Science, Massachusetts Institute of Technology, Cambridge, MA 02139, USA}

\corresau{Abigail Bodner, abodner@mit.edu}

\begin{document}
\maketitle

\begin{abstract}
Submesoscale currents in the oceanic mixed layer, comprising fronts, eddies, and filaments, are characterized by $\textit{O}(1)$ Rossby numbers ($\Ro$). These features, which constantly interact with background mesoscale flows and boundary layer turbulence (BLT), are critical for mediating vertical exchange between the surface and the ocean interior. Despite growing insight into their generation and evolution, the modification of initially balanced submesoscale dynamics by finite-$\Ro$ effects under the combined influence of mesoscale strain and BLT remains unresolved. In this study, we address this question through a perturbation analysis of two-dimensional, geostrophically adjusted oceanic fronts and filaments, adapting the analytical models of \citet{shakespeare_generalized_2013} and \citet{bodner_breakdown_2020}. This framework allows for a systematic exploration across a broad range of Rossby numbers $\Ro$, Ekman numbers ($\Ek$), and strain parameters. The first-order solution under pure mesoscale strain exhibits clear frontogenesis and closely mirrors the full model dynamics during early inertial periods, despite the absence of an exponential collapse. Under BLT perturbation, the first-order solution confirms the distinct frontogenetic and frontolytic tendencies associated with eddy viscosity and diffusivity, respectively; however, no transition between these regimes is observed across the explored $\Ro$ and $\Ek$ parameter space for vertical mixing. When both strain and BLT perturbations are present, turbulent fluxes can strengthen, weaken, or even reverse strain-induced frontogenesis depending on the parameter regime. These results suggest that mixed-layer parameterizations must carefully account for the spatial variability of BLT within submesoscale currents to accurately capture frontal evolution under mesoscale strain.
\end{abstract}

\begin{keywords}
Authors should not enter keywords on the manuscript, as these must be chosen by the author during the online submission process and will then be added during the typesetting process (see \href{https://www.cambridge.org/core/services/aop-file-manager/file/61436b61ff7f3cfab749ce3a/JFM-Keywords-Sept-2021.pdf.}{Keyword PDF} for the full list).  Other classifications will be added at the same time.
\end{keywords}


\section{Introduction}
\label{sec:intro}
Submesoscale flows in the upper ocean are dynamically rich and have significant socio-environmental implications. Characterized by horizontal scales of \qtyrange{0.1}{10}{\kilo\meter} and time scales of hours to days, they represent a critical dynamical bridge between the larger, quasi-geostrophic mesoscale and smaller, three-dimensional turbulence \citep[e.g.,][]{mcwilliams_submesoscale_2016,taylor_submesoscale_2023,ferrari_ocean_2009,molemaker_balanced_2010,callies_role_2016}. Their dynamics, where the effects of advection and planetary rotation are of comparable magnitude, are captured by an order-one Rossby number ($\Ro\sim 1$). This regime is marked by strong spatial and temporal inhomogeneity, which, while complicating analysis, gives rise to significant vertical velocities and sharp gradients in buoyancy and velocity near the surface \citep{capet_mesoscale_2008b,mahadevan_analysis_2006}. These intense vertical exchanges have profound practical implications, substantially contributing to the transport and dispersion of heat, biogeochemical tracers, and anthropogenic materials \citep{garrett_dynamical_1981,ferrari_frontal_2011,su_ocean_2018,mahadevan_impact_2016,taylor_accumulation_2018}. Despite their importance, these processes and their complex interactions with especially smaller-scale motions are not resolved in current global climate simulations. This resolution gap can introduce considerable uncertainty into our broader understanding of the ocean's role in the global climate system \citep[e.g.,][]{ferrari_frontal_2011}.

Among the most ubiquitous features of submesoscale features are fronts and filaments. Fronts are characterized by sharp horizontal gradients in properties like density in the cross-frontal direction, while exhibiting relative homogeneity along-front \citep{mcwilliams_oceanic_2021}. They are prevalent in regions of strong background vorticity and are seasonally more dominant in deep winter mixed layers \citep{callies_seasonality_2015}. Filaments can often be conceptualized as double-front structures separating warm or cold density anomalies from the surroundings and are similarly common in these dynamic environments \citep{mcwilliams_cold_2009}. The energy for the evolution of these structures is largely drawn from the background mesoscale strain field and from various forms of mixed-layer instabilities \citep[e.g.,][]{thomas_submesoscale_2008}. Concurrently, these submesoscale flows actively interact with smaller-scale boundary layer turbulence (BLT), modulating its structure and facilitating the eventual dissipation of energy \citep[e.g.,][]{hamlington_langmuir_2014, verma_interaction_2022, sullivan_frontogenesis_2018, callies_baroclinic_2018}.

The study of frontal dynamics has a rich history. Classic theories of frontogenesis mainly consider two dynamic mechanisms. One involves forcing by large-scale strain \citep{hoskins_atmospheric_1972}, while the other is related to geostrophic adjustment from an initially unbalanced state \citep{blumen_inertial_2000}. The former assumes a large time scale and ignores inertial oscillation, while the latter does not consider background strain. In a following work, \citet{shakespeare_generalized_2013} (hereafter ST13) unified these two mechanisms to consider both inertial and strain effects in one framework. Their theory permits comparable inertial and strain time scales and unbalanced initial conditions, more appropriate to submesoscale processes. All these pioneer works predict a finite-time singularity where the cross-frontal scale collapses exponentially. While foundational, this outcome is unphysical, as aforementioned instabilities and smaller-scale turbulence inevitably intervene. Although ST13 assumed that submesoscale strain rate could be comparable to the inertial frequency, their illustrations only involved small strain appropriate for mesoscale flows. And it is not uncommon to hold submesoscale processes with finite Rossby number in the oceanic mixed layer under mesoscale strain \citep{callies_role_2016}. 

Previous investigations have sought to characterize the influence of BLT on frontal evolution. Large Eddy Simulations (LES) have been particularly insightful, demonstrating that both vertical and horizontal turbulent fluxes are crucial in modulating frontogenesis \citep[e.g.,][]{sullivan_frontogenesis_2018,sullivan_oceanic_2024}. On the analytical front, a notable perturbation framework was developed by \citet{bodner_perturbation_2019} (hereafter B19), which serves as a conceptual basis for the present study. Their work introduced the effects of BLT into ST13 as a first-order correction to a zeroth-order, semi-geostrophic frontogenesis model, operating within a small Rossby number and small Ekman number asymptotic expansion. A key finding was the distinct roles of vertical and horizontal turbulence: vertical mixing tended to strengthen the front (frontogenesis), whereas horizontal mixing was largely frontolytic. While elegant, this framework is predicated on semi-geostrophic theory, which by design filters out ageostrophic inertia-gravity waves and assumes a small Rossby number, an assumption inconsistent with the $\Ro\sim 1$ submesoscale regime. Therefore, a theoretical framework that treats mesoscale strain and BLT as processes of the same perturbation order within a consistent submesoscale scaling remains less explored. 

In this study, we present a theoretical framework to investigate the evolution of oceanic fronts and filaments with finite Rossby numbers under the combined, comparable-order effects of mesoscale strain and BLT. This approach captures the time-dependent, ageostrophic responses of these submesoscale features by resolving their evolution within a momentum-following coordinate system. By doing so, we aim to illuminate the competitive interactions between large-scale forcing and small-scale mixing across a broad parameter space. We describe the theoretical framework in \S~\ref{sec:model} and formulate the asymptotic expansion to derive the closed set of equations governing the zeroth- and first-order solutions in \S~\ref{sec:perturbation}. In \S~\ref{sec:results}, we discuss results under a range of parameters followed by discussion and concluding remarks in \S~\ref{sec:discussion}.

\section{The model}
\label{sec:model}


The theoretical model builds on the ST13 and B19 theories for a strain-induced front oriented in the $y$ direction in the presence of turbulence. We consider an incompressible Boussinesq fluid on an $f$–plane subjected to a spatially uniform, steady horizontal strain–rate $\alpha>0$ that drives frontogenesis. The vertical domain is $0\le z\le H$ with the sea–surface at $z=H$ and the base of the seasonal mixed layer at $z=0$. 

The background straining flow is
\begin{equation}
  \bar U(x) = -\alpha x, \qquad \bar V(y) = \alpha y, \qquad
  \alpha>0.\label{eq:strain_flow}
\end{equation}
The velocity and pressure terms are written as
\begin{subequations}\label{eq:fields}
\begin{align}
  U &= \bar U(x)+u(x,z,t)=-\alpha x+u(x,z,t),\\
  V &= \bar V(x)+v(x,z,t)=\alpha y+v(x,z,t),\\
  W &= 0+w(x,z,t),\\
  P &= \bar P+p(x,z,t)=\rho_0\left[-\alpha^2\frac{(x^2+y^2)}{2}+f\alpha xy\right]+p(x,z,t),\\
  B &= 0+b(x,z,t),
\end{align}
\end{subequations}
where $\bar P$ is the pressure field associated with the background straining flow, $\rho_0$ is a reference density, and $u,v,w,p,b$ are the laminar frontogenetic velocity, pressure, and buoyancy fields.  

Following B19, we add $(x,z)$-dependent eddy viscosities and diffusivities representing the effects of BLT upon a hydrostatic, laminar flow, yielding the following governing equations,
\begin{subequations}\label{eq:dimensional}
\begin{align}
  \frac{\D u}{\D t} - f v &= \alpha u - \frac{1}{\rho_0}\frac{\p p}{\p x} + \frac{\p}{\p x} \left(\nu_H\frac{\p u}{\p x}\right) + \frac{\p}{\p z}\left(\nu_V \frac{\p u}{\p z}\right),\\
  \frac{\D v}{\D t} + f u &= -\alpha v  + \frac{\p}{\p x} \left(\nu_H\frac{\p v}{\p x}\right) + \frac{\p}{\p z}\left(\nu_V \frac{\p v}{\p z}\right),\\
  0 &= b - \frac{1}{\rho_0}\frac{\p p}{\p z},\\
  \frac{\D b}{\D t} &= \frac{\p}{\p x} \left(\kappa_H\frac{\p b}{\p x}\right) + \frac{\p}{\p z}\left(\kappa_V \frac{\p b}{\p z}\right),\\
  \frac{\p u}{\p x} + \frac{\p w}{\p z}&= 0,
\end{align}
\end{subequations}
with the material derivative given by,
\begin{equation}
    \frac{\D }{\D t} \equiv \frac{\p}{\p t} + (u + \bar U)\frac{\p}{\p x} + w \frac{\p}{\p z}.
\end{equation}
These are the same equations as in B19. We follow ST13 to consider zero initial cross-front  flow $u=w=0$ and an along-front velocity of $v=v_g$, where $v_g$ is a geostrophic velocity defined by the pressure field and in balance with the buoyancy field as
\begin{equation}\label{eq:defvg}
    \frac{\p v_g}{\p z}=\frac{\p}{\p z}\left(\frac{1}{\rho_0 f}\frac{\p p}{\p x}\right)= \frac{1}{f}\frac{\p b}{\p x}.
\end{equation}
The initial $b$ also follows the one in ST13 to be defined in a dimensionless form below.

\subsection{Dimensionless expressions}
\label{sec:nondim}
Following B19, we use  scaling parameters to obtain dimensionless perturbation equations: the horizontal and vertical buoyancy gradients ($M^2\sim\p b/\p y$ and $N^2\sim\p b/\p z$, which is also the buoyancy frequency squared), the horizontal and vertical length scales $(L, H)$, the strain and Coriolis rate parameters $(\alpha, f )$, and scales for viscosities and diffusivities $(\nu_H,\nu_V,\kappa_H,\kappa_V)$. The dimensionless expressions for quantities of interest are given in table~\ref{tab:symbols}. As in B19, the vertical dimensionless coordinate ranges from 0 to 1, 0 being the bottom of the mixed layer and 1 the surface. The cross-frontal coordinate is centred around the initial front maximum. 

The dimensionless versions of \eqref{eq:dimensional} are, after reorganizing,
\begin{subequations}\label{eq:dimensionless}
\begin{align}
  \frac{\D u}{\D t} -  \Ro\ v  -{\gamma} u&=  - \Ro\frac{\p p}{\p x} + {\Ek_H}\frac{\p}{\p x} \left(\nu_H\frac{\p u}{\p x}\right) + {\Ek_V}\frac{\p}{\p z}\left(\nu_V \frac{\p u}{\p z}\right),\\
  {\frac{\D v}{\D t}} + \frac{1}{\Ro} u  &= -{\gamma}v + {\Ek_H}\frac{\p}{\p x} \left(\nu_H\frac{\p v}{\p x}\right) + {\Ek_V}\frac{\p}{\p z}\left(\nu_V \frac{\p v}{\p z}\right),\\
    b&= \frac{\p p}{\p z},\\
  {\frac{\D b}{\D t} } &= \frac{\Ek_H}{\Pran_H}\frac{\p}{\p x} \left(\kappa_H\frac{\p b}{\p x}\right) + \frac{\Ek_V}{ \Pran_V}\frac{\p}{\p z}\left(\kappa_V \frac{\p b}{\p z}\right),\\
  \frac{\p u}{\p x}  + \frac{\p w}{\p z}  &= 0,
\end{align}
\end{subequations}
where
\begin{equation}
    \frac{\D}{\D t}=\frac{\p}{\p t}+\left({\Ro}\ u-{\gamma} x\right)\frac{\p}{\p x}+ {\Ro}\ w \frac{\p}{\p z}.
\end{equation}
The dimensionless along-front momentum and buoyancy equations are intended to be identical to those in B19 once we assume constant viscosities and diffusivities. The initial velocity condition is $u=w=0$ and $v=v_g$. For the initial buoyancy, as in ST13, we suppose an initially zero potential vorticity with 
\begin{equation}
  b(x,z,0) = B_0(x+\Ro^2 v),
\end{equation}
where the buoyancy anomaly $\Delta b$ vanishes at $z=0,1$. Note that a factor of $\Ro$ is added compared to the ST13 formulation due to different along-front velocity scales. Boundary conditions are flat, and we do not consider explicit surface forcing effects beyond BLT terms. A useful diagnostic for frontogenesis is the frontal width $d$ defined in ST13 as
\begin{equation}
    d=\frac{\max{\left|\frac{\p B_0}{\p X}\right|}}{\max{\left|\frac{\p b}{\p x}\right|}}.
\end{equation}

Note that the system consists of different characteristic Rossby numbers for the cross-front flow $\Ro=\sqrt{M^2H^3/L}$, the along-front flow $\Ro^2M^2H^3/L$, and the deformation ratio  of the large-scale strain flow $\gamma=\alpha/f$. Building on the framework of B19, here we present a more general perturbation approach with $\Ro=\textit{O}(1)$, as is appropriate for submesoscale dynamics. Furthermore, we take advantage of the generalized model by ST13, and now assume the baseline solution of a front under geostrophic balance, on top of which both strain and BLT affect and compete at the next order. 

\begin{table}
  \begin{center}
\def~{\hphantom{0}}
  \begin{tabular}{lcc}
          & Symbol   &   Scale \\[3pt]
       Buoyancy   & $b$ & $M^2L$\\[3pt]
       Geostrophic velocity   & $v$ & $\frac{HM^2}{f}$\\[3pt]
       Horizontal velocity  & $u$ & $\sqrt{M^2LH}$\\[3pt]
       Vertical velocity   & $w$ & $\sqrt{\frac{M^2H^3}{L}}$\\[3pt]
       Time & t & $\frac{1}{f}$\\[3pt]
       Background mean velocity & $\bar{U}$        & $\alpha\,L$ \\[3pt]
    Rossby number & $\Ro$                             & $\sqrt{\frac{M^{2}H}{f^{2}L}}$ \\[3pt]
    Deformation ratio& $\gamma$                     & $\frac{\alpha}{f}$ \\[3pt]
    Horizontal Ekman number & $\Ek_H$ & $\frac{\bar \nu_H}{fL^2}$ \\[3pt]
    Vertical Ekman number & $\Ek_V$ & $\frac{\bar \nu_V}{fH^2}$ \\[3pt]
    Horizontal Prandtl number& $\Pran_H$       & $\frac{\bar \nu_H}{\bar \kappa_H}$ \\[3pt]
    Vertical Prandtl number& $\Pran_V$       & $\frac{\bar \nu_V}{\bar \kappa_V}$ \\[3pt]
  \end{tabular}
  \caption{Dimensionless expressions for quantities of interest following B19 framework.}
  \label{tab:symbols}
  \end{center}
\end{table}

\section{Perturbation analysis with \textit{O}(1) Rossby number}
\label{sec:perturbation}

We propose a perturbation approach with a small term $\varepsilon$ accounting for higher order effects due to strain and BLT, thus allowing them to balance and compete with one another. We construct the zeroth-order solution to be a laminar frontal dynamics in geostrophic balance described in ST13 that permits \textit{O}(1) Rossby number.
\begin{subequations}\label{eq:perturbation}
\begin{align}
  U &= \bar U+u=\varepsilon^0(\bar U+u^{0})+\varepsilon^1 u^{1}+\textit{O}(\varepsilon^2),\\
  V &= \bar V+v=\varepsilon^0(\bar V+v^{0})+\varepsilon^1 v^{1}+\textit{O}(\varepsilon^2),\\
  W &= w=\varepsilon^0 w^{0}+\varepsilon^1 w^{1}+\textit{O}(\varepsilon^2),\\
  B &= b=\varepsilon^0 b^{0}+\varepsilon^1 b^{1}+\textit{O}(\varepsilon^2).
\end{align}
\end{subequations}
Assuming $\Ro=\textit{O}(1)$ for submesoscale dynamics, for tractability, we define separate perturbation parameters for each of the higher order forcing terms, representative of strain, and horizontal and vertical viscosity and diffusivity,
\begin{equation}\label{eq:perturbscale}
\varepsilon_S\equiv \gamma,\ \varepsilon_{HV} \equiv {\Ek_H},\ \varepsilon_{VV} \equiv {\Ek_V},\ \varepsilon_{HD} \equiv \frac{\Ek_H}{\Pran_H},\ \varepsilon_{VD} \equiv \frac{\Ek_V}{\Pran_V}.
\end{equation}
The calculation example in Appendix~\ref{app:scaleexp} demonstrates that the perturbation assumption is reasonable for typical values and consistent with $\varepsilon<1$. We explore below the general parameter space by treating terms in \eqref{eq:perturbscale} as of the same order. 

\subsection{Zeroth–order geostrophic balance} \label{subsec:zeroth}
At $\textit{O}(\varepsilon^0)$ we obtain
\begin{subequations}\label{eq:zeroth}
\begin{align}
  \left(\frac{1}{\Ro}\frac{\p }{\p t}+u^0\frac{\p }{\p x}+ w^0 \frac{\p}{\p z}\right) u^0 -  v^0 &=  - \frac{\p p^0}{\p x} ,\\
  \left(\frac{1}{\Ro}\frac{\p }{\p t}+u^0\frac{\p }{\p x}+ w^0 \frac{\p}{\p z}\right) v^0  &= - \frac{1}{\Ro^2} u^0 ,\\
    0 &= b^0 - \frac{\p p^0}{\p z},\\
  \left(\frac{1}{\Ro}\frac{\p }{\p t}+u^0\frac{\p }{\p x}+ w^0 \frac{\p}{\p z}\right) b^0&= 0,\\
  \frac{\p u^0}{\p x}  + \frac{\p w^0}{\p z}  &= 0,
\end{align}
\end{subequations}
with 
initial conditions $u^0=w^0=0$, $v^0=v_g$, and $b^0=B_0$. An approximate solution $(u^{0}, v^{0}, w^{0},b^{0})$, assuming zero potential vorticity (PV) and neglecting higher-order nonlinear terms, is given by setting $\delta=0$ in equations (3.4) of ST13, corresponds to geostrophic balance that permits $\Ro=\textit{O}(1)$. In comparison, the zeroth order of B19 is the longer-term semi-geostrophic approximation with no inertial oscillations. Here we reproduce the solution in ST13 as a zeroth-order approximation.

We adapt the zeroth-order momentum coordinates as in ST13 for the geostrophic adjustment case without strain
\begin{equation}
  X_0 =  (x+\Ro^2 v^0), \qquad Z = z, \qquad T = t.
\end{equation}
Note that a factor of $\Ro$ is added compared to the ST13 formulation due to different along-front velocity scales. In these variables the $X_0$ coordinate is conserved, and given an arbitrary lateral buoyancy profile $B_0(X_0)$, the buoyancy profile is fixed.
In ST13, $B_0(X_0)$ is chosen to be $B_0(X_0)=\frac{1}{2}\mathrm{erf}(X_0/\sqrt{2})$ for a front. Here we also consider a cold submesoscale filament with $B_0(X_0)=\frac{1}{2}\exp(-X_0^2/2)$ \citep[e.g.,][]{sullivan_frontogenesis_2018}. 
In terms of velocity components, from ST13 (3.4) with negligible strain we have
\begin{subequations}\label{eq:vel0}
\begin{align}
  u^{0}(X_0,Z,T) &= 0,\\[4pt]
  v^{0}(X_0,Z,T) &= \,B_0'(X_0)\left(Z-\frac12\right),\\[4pt]
  w^{0}(X_0,Z,T) &= 0.
\end{align}
\end{subequations}

Note that the zeroth-order inverse Jacobian of the coordinate transformation is
\begin{equation}
    J_0^{-1}=1-\Ro^2 \frac{\p v^0}{\p X_0}.
\end{equation}
which is time-independent in this model. Therefore, the perturbation approach with a zeroth-order geostrophic balance holds only if $J_0^{-1}>0$, which leads to a constraint on the Rossby number as
\begin{equation}
    \Ro < \Ro_c = \sqrt{\frac{2}{\max{|B''_0(X_0)|}}}.
\end{equation}
With $B_0(X_0)=\frac{1}{2}\mathrm{erf}(X_0/\sqrt{2})$, we have $\Ro_c=2.92$, so $\Ro=\textit{O}(1)$ is achievable. The filament case has a lower $\Ro_c=2$ but still greater than 1. The zeroth-order expressions will be treated as known coefficients in the first-order system. We will see that the cleanness of the geostrophic balance permits the use of a single prognostic variable in this system, which is the streamfunction of the ageostrophic secondary circulation. The first-order solution of this variable reflects the combined effects of background strain and BLT. 


\subsection{First–order system with strain and turbulence} 
The governing equations of the first-order system are
\begin{subequations}\label{eq:first}
\begin{align}
  \frac{\D u^1}{\D t_0} +\left(\vec{U^1 \cdot \nabla}\right)u^0- \Ro\left( v^1 - \frac{\p p^1}{\p x}\right)&=  {\widetilde \varepsilon_S} u^0  + {\widetilde \varepsilon_{HV}} \frac{\p}{\p x} \left(\nu_H\frac{\p u}{\p x}\right)+ {\widetilde \varepsilon_{VV}} \frac{\p}{\p z}\left(\nu_V \frac{\p u^0}{\p z}\right),\\
  \frac{\D v^1}{\D t_0} +\left(\vec{U^1 \cdot \nabla}\right)v^0 + \frac{1}{\Ro} u^1  &= -{\widetilde \varepsilon_{S}} v^0 + {\widetilde \varepsilon_{HV}}\frac{\p}{\p x} \left(\nu_H\frac{\p v}{\p x}\right)+ {\widetilde \varepsilon_{VV}}\frac{\p}{\p z}\left(\nu_V \frac{\p v^0}{\p z}\right),\\
    0 &= b^1 - \frac{\p p^1}{\p z},\\
  \frac{\D b^1}{\D t_0} +\left(\vec{U^1 \cdot \nabla}\right)b^0 &= {\widetilde{\varepsilon}_{HD}}\frac{\p}{\p x}\left(\kappa_H \frac{\p b^0}{\p x}\right)+{\widetilde{\varepsilon}_{VD}}\frac{\p}{\p z}\left(\kappa_V \frac{\p b^0}{\p z}\right),\\
  \frac{\p u^1}{\p x}  + \frac{\p w^1}{\p z}  &= 0,
\end{align}
\end{subequations}
where 
\begin{equation}
    \frac{\D}{\D t_0}=\frac{\p}{\p t}+{\Ro}\left(u^0\frac{\p}{\p x}+ w^0 \frac{\p}{\p z}\right),\quad \vec{U^1 \cdot \nabla}=\left({\Ro}\ u^1-{\widetilde \varepsilon_S}x\right)\frac{\p}{\p x}+{\Ro}\ w^1\frac{\p}{\p z},
\end{equation}
$\widetilde{\varepsilon_i}= \varepsilon^{-1}{\varepsilon_i}$, with $i\in\{S,HV,VV,HD,VD\}$ and $\widetilde{\varepsilon_i}=1$ for at least one of them as the perturbation parameter. 
The initial condition is $u^1=v^1=w^1=b^1=0$. Note that we do not assume $v^1$ and $b^1$ are in geostrophic balance. This means that the first order effects have the ability to impose ageostrophic effects on the front. 
An analytical solution to this system is unlikely due to the presence of general, non-constant coefficient fields $(u^0,w^0,\nu_V,\kappa_V)$ that depend on both space and time. Such complexity prevents the use of standard analytical techniques like separation of variables or integral transforms in their usual form. 

Nevertheless, given the zeroth-order geostrophic balance in \S~\ref{subsec:zeroth}, the governing linear system \eqref{eq:first} simplifies considerably. The explicit spatial dependence of the background strain is confined to forcing terms and does not appear in the differential operators acting on the first-order unknowns. This allows for a more straightforward analysis. First, we reformulate the governing equations in terms of three unknowns. To satisfy the continuity equation and eliminate pressure, we introduce a zeroth-order streamfunction $\phi^0$ and a first-order streamfunction $\psi^1$ such that  
\begin{equation}
    v^0 = \frac{\p \phi^0}{\p x},\quad b^0=\frac{\p \phi^0}{\p z}, \quad u^1 = \frac{\p \psi^1}{\p z},\quad w^1=-\frac{\p \psi^1}{\p x}.
\end{equation}
With the forcing terms $F_v$ and $F_b$ given by
\begin{align}\label{eq:fvfb}
    F_v &= -\widetilde{\varepsilon}_S \frac{\p \phi^0}{\p x} + \widetilde{\varepsilon}_{HV} \frac{\p}{\p x}\left(\nu_H \frac{\p^2 \phi^0}{\p x^2 }\right)+  \widetilde{\varepsilon}_{VV}\frac{\p}{\p z}\left(\nu_V \frac{\p^2 \phi^0}{\p x\p z}\right) + \widetilde{\varepsilon}_S x\frac{\p^2 \phi^0}{\p x^2}, \\
    F_b &=  \widetilde{\varepsilon}_{HD}\frac{\p}{\p x}\left(\kappa_H \frac{\p^2 \phi^0}{\p x\p z}\right)+ \widetilde{\varepsilon}_{VD}\frac{\p}{\p z}\left(\kappa_V \frac{\p^2 \phi^0}{\p z^2}\right) +\widetilde{\varepsilon}_S x\frac{\p^2 \phi^0}{\p x\p z},
\end{align}
pressure is given by the hydrostatic relationship to obtain a system for $(\psi^1, v^1, b^1)$: 
\begin{subequations} \label{eq:psi_system}
\begin{align}
    \frac{1}{\Ro}\frac{\p}{\p t}\left(\frac{\p^2 \psi^1}{\p z^2}\right) - \frac{\p v^1}{\p z} + \frac{\p b^1}{\p x} &= 0, \\
    \frac{\p v^1}{\p t} + \left(\frac{1}{\Ro}+{\Ro}\frac{\p^2 \phi^0}{\p x^2}\right)\frac{\p \psi^1}{\p z}-{\Ro}\frac{\p^2 \phi^0}{\p x\p z}\frac{\p \psi^1}{\p x} &= {F}_v(x,z; \phi^0), \\
    \frac{\p b^1}{\p t} +{\Ro}\left(\frac{\p^2 \phi^0}{\p x\p z}\frac{\p \psi^1}{\p z}-\frac{\p^2 \phi^0}{\p z^2}\frac{\p \psi^1}{\p x}\right)&= {F}_b(x,z; \phi^0),
\end{align}
\end{subequations}
We further reduce the system to a single equation for $\psi^1$ as
\begin{equation} \label{eq:psi1}
    \frac{1}{\Ro}\frac{\p^4 \psi^1}{\p t^2 \p z^2}+ {\Ro}\left[\left(\frac{1}{\Ro^2}+\frac{\p^2 \phi^0}{\p x^2}\right)\frac{\p^2 \psi^1}{\p z^2}-2\frac{\p^2 \phi^0}{\p x\p z}\frac{\p^2 \psi^1}{\p x\p z}+\frac{\p^2 \phi^0}{\p z^2}\frac{\p^2 \psi^1}{\p x^2}\right]=\frac{\p F_v}{\p z}-\frac{\p F_b}{\p x}.
\end{equation}
This governing equation for the streamfunction perturbation $\psi^1(x,z,t)$ is a linear, inhomogeneous partial differential equation. Substituting the coefficients derived from $v^0$ and $b^0$ for $\phi^0$, the equation takes the form
\begin{equation}
\frac{\partial^2}{\partial t^2}\left(\frac{\partial^2\psi^1}{\partial z^2}\right) + J_0 \left( \frac{\partial}{\partial z} -\Ro^2\, B'_0 \frac{\partial}{\partial x} \right)^2 \psi^1 = F_1(x,z)\equiv\Ro \, \left(\frac{\p F_v}{\p z}-\frac{\p F_b}{\p x}\right),
\label{eq:gov_final}
\end{equation}
The solution is sought subject to the boundary conditions $\psi^1(z=0,1)=0$ and initial conditions $\psi^1(t=0)=0$ and $\psi^1_t(t=0)=0$. 

The most direct method of solution is to employ superposition, decomposing the streamfunction into a time-independent particular solution, $\psi^1_p(x,z)$, and a time-dependent homogeneous solution, $\psi^1_h(x,z,t)$:
\begin{equation}
\psi^1(x,z,t) = \psi^1_p(x,z) + \psi^1_h(x,z,t).
\end{equation}
The particular solution $\psi^1_p$ satisfies the steady-state equation, balancing the background forcing:
\begin{equation}
J_0 \left( \frac{\partial}{\partial z} - \Ro^2\,B'_0 \frac{\partial}{\partial x} \right)^2 \psi^1_p = F_1(x,z).
\label{eq:particular_eqn}
\end{equation}
This equation is simplified by a change of coordinates from $(x,z)$ to $(X_0,Z)$. In this new frame, the advective derivative operator becomes $\frac{\partial}{\partial z} - \Ro^2\, B'_0 \frac{\partial}{\partial x} = \frac{\partial}{\partial Z}$, and the spatial derivative transforms as $\frac{\partial}{\partial x} = J_0\frac{\partial}{\partial X}$. Applying this transformation to (\ref{eq:particular_eqn}) yields
\begin{equation}
\frac{\partial^2 \tilde{\psi}^1_p}{\partial Z^2} =J_0^{-1} F_1(X_0,Z),
\label{eq:particular_ode}
\end{equation}
where tildes denote functions in the $(X_0,Z)$ coordinate system. 
Equation (\ref{eq:particular_ode}) is an ordinary differential equation for $\tilde{\psi}^1_p(X_0,Z)$ in the variable $Z$, with $X_0$ treated as a parameter. We solve (\ref{eq:particular_ode}) by direct integration:
\begin{equation}
\tilde{\psi}^1_p(X_0,Z) = \int_0^Z \int_0^{Z'} J_0^{-1} F_1(X_0,Z'')\,\mathrm{d}Z''\,\mathrm{d}Z' + C_1(X_0)Z + C_2(X_0).
\end{equation}
The functions $C_1(X_0)$ and $C_2(X_0)$ are determined by the boundary conditions $\tilde{\psi}^1_p(X_0,0)=0$ and $\tilde{\psi}^1_p(X_0,1)=0$, yielding $C_2(X_0)=0$ and $C_1(X_0) = -\int_0^1 \int_0^{Z'} J_0^{-1}F_1(X_0,Z'')\,\mathrm{d}Z''\,\mathrm{d}Z'$. This procedure fully specifies the particular solution $\psi^1_p$.

The homogeneous solution $\psi^1_h$ satisfies the unforced equation
\begin{equation}
\frac{\partial^2}{\partial t^2}\left(\frac{\partial^2\psi^1_h}{\partial z^2}\right) + J_0 \left( \frac{\partial}{\partial z} - \Ro^2\, B'_0 \frac{\partial}{\partial x} \right)^2 \psi^1_h = 0,
\label{eq:homogeneous_eqn}
\end{equation}
with homogeneous boundary conditions and modified initial conditions derived from the total solution:
\begin{equation}
\psi^1_h(t=0) = -\psi^1_p(x,z), \quad \frac{\partial\psi^1_h}{\partial t}(t=0) = 0.
\end{equation}

We solve (\ref{eq:homogeneous_eqn}) via a normal mode analysis, seeking solutions of the form $\psi^1_h(x,z,t) = \text{Re}[\hat{\psi}^1(x,z)e^{-i\omega t}]$. This leads to a generalized eigenvalue problem for the spatial mode structure $\hat{\psi}^1$ and the frequency $\omega$:
\begin{equation}
\mathcal{L}[\hat{\psi}^1] \equiv J_0 \left( \frac{\partial}{\partial z} -\Ro^2\, B'_0 \frac{\partial}{\partial x} \right)^2 \hat{\psi}^1 = \omega^2 \frac{\partial^2\hat{\psi}^1}{\partial z^2}.
\label{eq:eigen_problem}
\end{equation}
Solving (\ref{eq:eigen_problem}) provides a set of eigenfunctions $\hat{\psi}^1_n$ and corresponding eigenfrequencies $\omega_n$. The general homogeneous solution is a linear combination of these modes. Given the initial condition $\psi^1_{h,t}(t=0)=0$, the solution simplifies to a cosine series:
\begin{equation}
\psi^1_h(x,z,t) = \sum_n A_n \cos(\omega_n t) \hat{\psi}^1_n(x,z).
\end{equation}
The coefficients $A_n$ are found by projecting the initial condition $-\psi^1_p$ onto the eigenbasis:
\begin{equation}
\sum_n A_n \hat{\psi}^1_n(x,z) = -\psi^1_p(x,z).
\end{equation}
Assuming the eigenfunctions form a complete set, orthogonal under an appropriate inner product, the coefficients $A_n$ are uniquely determined. Dynamically, the unforced first-order solution with a zeroth-order geostrophic balance is equivalent to the evolution of a balanced state under perturbed higher-order imbalance. And this leads to the geostrophic adjustment solution in ST13. Therefore, we argue that the homogeneous solution is dominated at the inertial frequency $\omega=1$ so that $\psi^1_h(x,z,t) \approx -\psi^1_p(x,z) \cos{t}$. Although this is not an exact solution of \eqref{eq:homogeneous_eqn}, we demonstrate later that it only has higher-order errors near boundaries and large gradients, so this approximation is consistent with the perturbation analysis.


The final solution is then the sum $\psi^1 = \psi^1_p + \psi^1_h$, which can be approximated as $\psi^1 \approx (1-\cos{t})\psi^1_p$, assuming the homogeneous component is dominated by near-inertial modes. Given $\psi^1 \approx (1-\cos{t})\psi^1_p$, the corresponding solutions for $v^1$ and $b^1$ are then 
\begin{subequations} \label{eq:v1b1_blt}
\begin{align}
    {v^1} &\approx {F}_v(x,z; \phi^0)t-\Ro\left[ \left(\frac{1}{\Ro^2}+\frac{\p^2 \phi^0}{\p x^2}\right)\frac{\p \psi^1_p}{\p z}-\frac{\p^2 \phi^0}{\p x\p z}\frac{\p \psi^1_p}{\p x}\right]\left(t-\sin{t}\right), \\
    {b^1} &\approx {F}_b(x,z; \phi^0)t-\Ro\left(\frac{\p^2 \phi^0}{\p x\p z}\frac{\p \psi_p^1}{\p z}-\frac{\p^2 \phi^0}{\p z^2}\frac{\p \psi_p^1}{\p x}\right)\left(t-\sin{t}\right).
\end{align}
\end{subequations}
We note that the time dependence in the first order system consists of a linear and sinusoidal part. This defers from the classic frontogenesis case as it does not include an exponential term. Another useful diagnostic measure for frontal tendency is the Lagrangian evolution of the horizontal buoyancy gradient \citep{bodner_perturbation_2019}
\begin{equation}
    T_b=\frac{\D}{\D t}\frac{1}{2}\left(\frac{\p b}{\p x}\right)^2=T_b^0+\epsilon T_b^1+\textit{O}(\epsilon^2),
\end{equation}
with $T_b^0=0$ due to zeroth-order geostrophic balance and
\begin{equation}
    T_b^1=\frac{1}{\Ro}\frac{\p^2 \phi^0}{\p x\p z} \frac{\p^2 b^1}{\p t \p x}+\frac{1}{2}\left[\left(\frac{\p \psi^1}{\p z}-\frac{\widetilde \varepsilon_S}{\Ro}x\right)\frac{\p}{\p x}-\frac{\p \psi^1}{\p x}\frac{\p}{\p z}\right]\left(\frac{\p^2 \phi^0}{\p x\p z}\right)^2,
\end{equation}
which consists of a constant and sinusoidal parts. Therefore, the first-order model describes a quasi-linear frontal strengthening or weakening under comparable strain and BLT perturbations. In particular, as we split $F_1$ into different mechanisms as
\begin{equation}
    F_1=F_S+F_{HV}+F_{VV}+F_{HD}+F_{VD},
\end{equation}
which consists of contributions from strain ($F_S$) and BLT ($F_{HV},F_{VV},F_{HD},F_{VD}$). We first examine them separately and then explore their combined effect.

\section{Results}\label{sec:results}
\subsection{Zeroth-order solution at $\Ro=1$}
\begin{figure}[t]
  \centerline{\includegraphics[width=\linewidth]{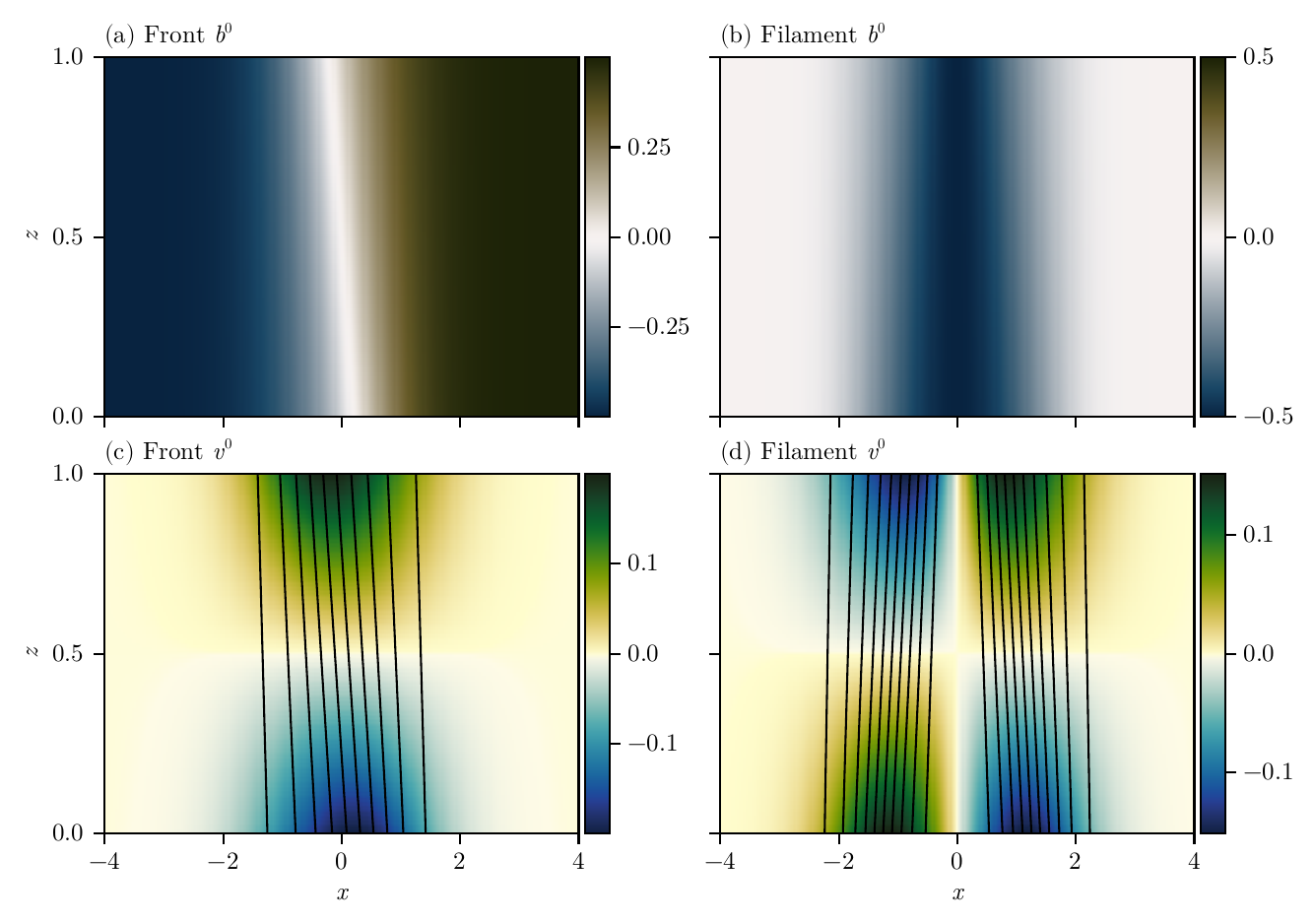}}
  \caption{A submesoscale front and filament in geostrophic balance. They correspond to the zeroth order solution $b^0=B_0(X_0)$ and $v^{0} = \,B_0'(X_0)(Z-1/2)$ with $\Ro=1$, $X_0 =  (x+\Ro^2 v^0)$, and $Z = z$.}
\label{fig:gb1}
\end{figure}




In the following sections, we examine solutions to specific parameter choices within the perturbation  framework. With $\Ro=1$, we have a steady geostrophic balance of a submesoscale front and filament (figure~\ref{fig:gb1}). Note that the front buoyancy gradient is slightly tilted in the original $(x,z)$ coordinate due to the adjustment, and that for the filament is symmetrically tilted towards the center at the top surface.

\subsection{First-order solution: strain-only}
In the strain-only case, we set ${\widetilde \varepsilon_S}=1$ and all BLT terms to zero, giving 
\begin{equation}
    F_S={\Ro}\left[\frac{\partial}{\partial z}\left(- \frac{\p \phi^0}{\p x}+x\frac{\p^2 \phi^0}{\p x^2}\right)-\frac{\partial}{\partial x}\left(x\frac{\p^2 \phi^0}{\p x\p z}\right)\right]=-2{\Ro} B'_0 J_0,
\end{equation}
and the corresponding particular solution is
\begin{equation}\label{eq:psi1p}
    \psi^1_{p,S}=-\Ro \,{B_0'(X_0)}Z(Z-1).
\end{equation}
We can obtain an approximate full solution with $\psi^1_{S}\approx (1-\cos{T})\psi^1_{p,S}$. The first-order velocity and buoyancy can be derived from $\psi^1_{S}$ as
\begin{subequations}\label{eq:v1b1strain}
\begin{align}
    v^1_S&=\left(Z-\frac12\right)\frac{T[B''_0(X_0)X_0+B'_0(X_0)]-2 {B'_0(X_0)}\sin{T}}{1-\Ro^2\, B''_0(X_0)(Z-\frac12)},\\
    b^1_S&= {B'_0(X_0)}(TX_0+\Ro^2\, v^1_S).
\end{align}
\end{subequations}
We show the derivation details, comparison with ST13, and analysis of the approximation accuracy in Appendix~\ref{app:strainerror}.

\begin{figure}
  \centerline{\includegraphics[width=\linewidth]{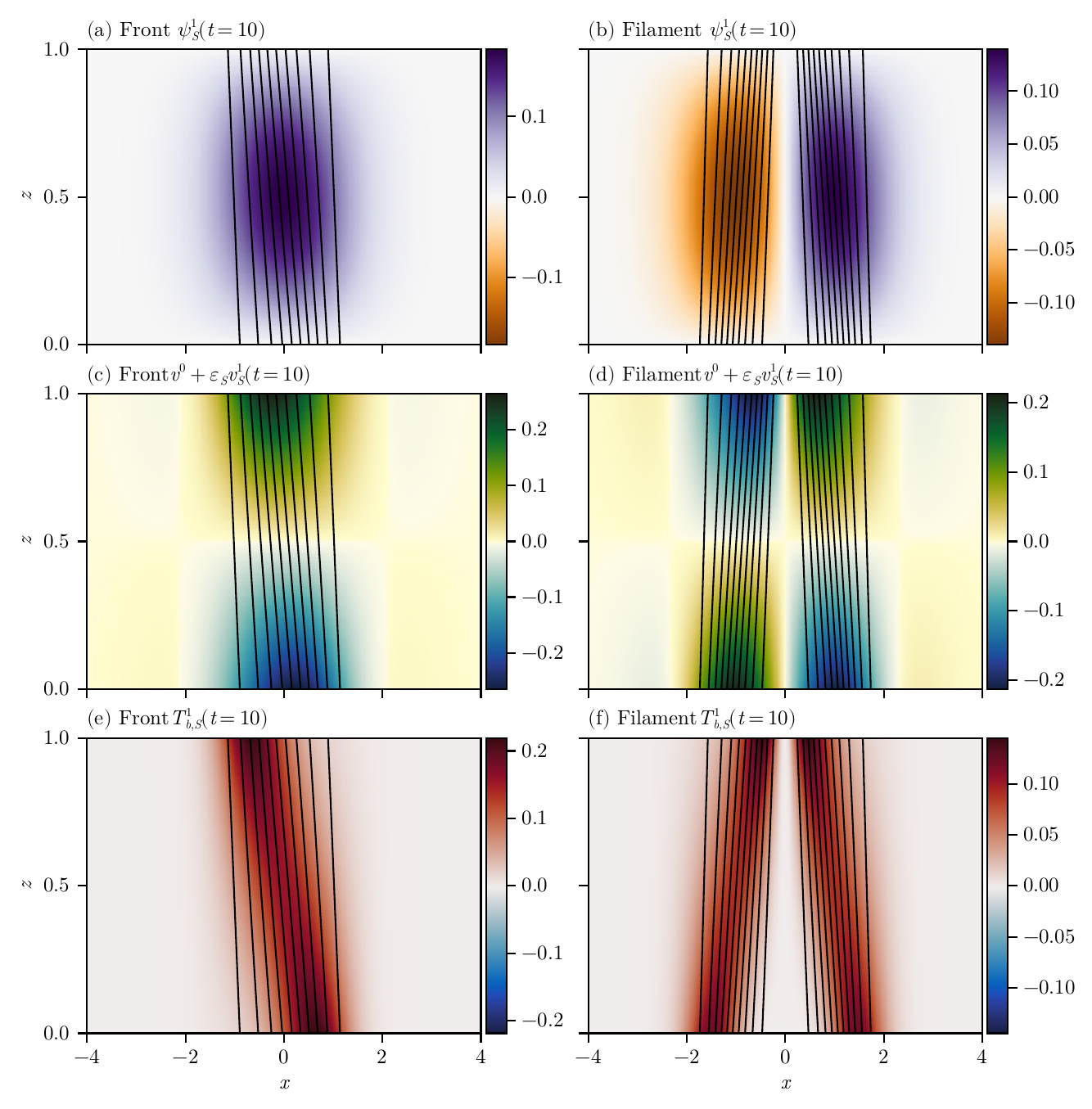}}
  \caption{Front and filament profiles of (a, b)  Streamfunction solutions $\psi^1_{S}$, (c, d) Velocity solutions $v^0+\varepsilon_S v^1_S$, and (e, f) First-order frontal tendency $T^1_{b,S}$. Black contours show the buoyancy $b^0+\varepsilon_S b^1_S$. The strain parameter is $\varepsilon_S=0.03$, and the corresponding time is $t=10$.}
\label{fig:strainff}
\end{figure}
Here we provide an example for a case of $\Ro=1$ and $\varepsilon_S=0.03$, a typical mesoscale strain, at $t=10$. Figure~\ref{fig:strainff}ab shows the streamfunction $\psi^1_{S}$ for front and filament. Both solutions indicate secondary circulations that flow towards colder (warmer) fluid near the surface (bottom), acting to flatten the front and filament as expected. The full secondary circulation has the same fixed spatial pattern but oscillates in time, instead of concentrating exponentially as in ST13 or B19. Figure~\ref{fig:strainff}cd show concentrating velocity profiles $v^0+\varepsilon_S v^1_S$ and buoyancy contours $b^0+\varepsilon_S b^1_S$. While buoyancy forcings $F_{b,S}$ have a strengthening effect, velocity forcings $F_{v,S}$ oppose the zeroth-order solution away from the center, indirectly sharpening the front (figure~\ref{fig:fbfv}). Figure~\ref{fig:strainff}ef shows the corresponding first-order frontal tendency indicator $T^1_{b,S}$ that is consistently strengthened at top and bottom boundaries, where $\psi^1_{p,S}$ indicates convergence.

\subsection{First-order solution induced by boundary-layer turbulence}
For simplicity, we here assume $\Ro=1$ with constant viscosities and diffusivities. Yet note that the analysis can be applied to more generalized cases. Assuming $\Pran_V<1$, $\Pran_H<1$ and $\nu_V=\nu_H=\kappa_V=\kappa_H=1$ after dimensionless normalization, typical of BLT in the mixed layer \citep[e.g.,][]{sullivan_frontogenesis_2018}, the effect of BLT on $\psi^1$ comes from the vertical and horizontal diffusion fluxes as
\begin{equation}\label{eq:fvhdv}
    F_{VD}\propto-\frac{\p^4 \phi^0}{\p x\p z^3},\quad F_{VV} \propto -F_{VD},\quad F_{HD}\propto -\frac{\p^4 \phi^0}{\p x^3\p z}, \quad F_{HV}\propto -F_{HD},
\end{equation}
which arrives from combining \eqref{eq:fvfb} with the definition of $F_1$ in \eqref{eq:gov_final} for each case.

\begin{figure}
  \centerline{\includegraphics[width=\linewidth]{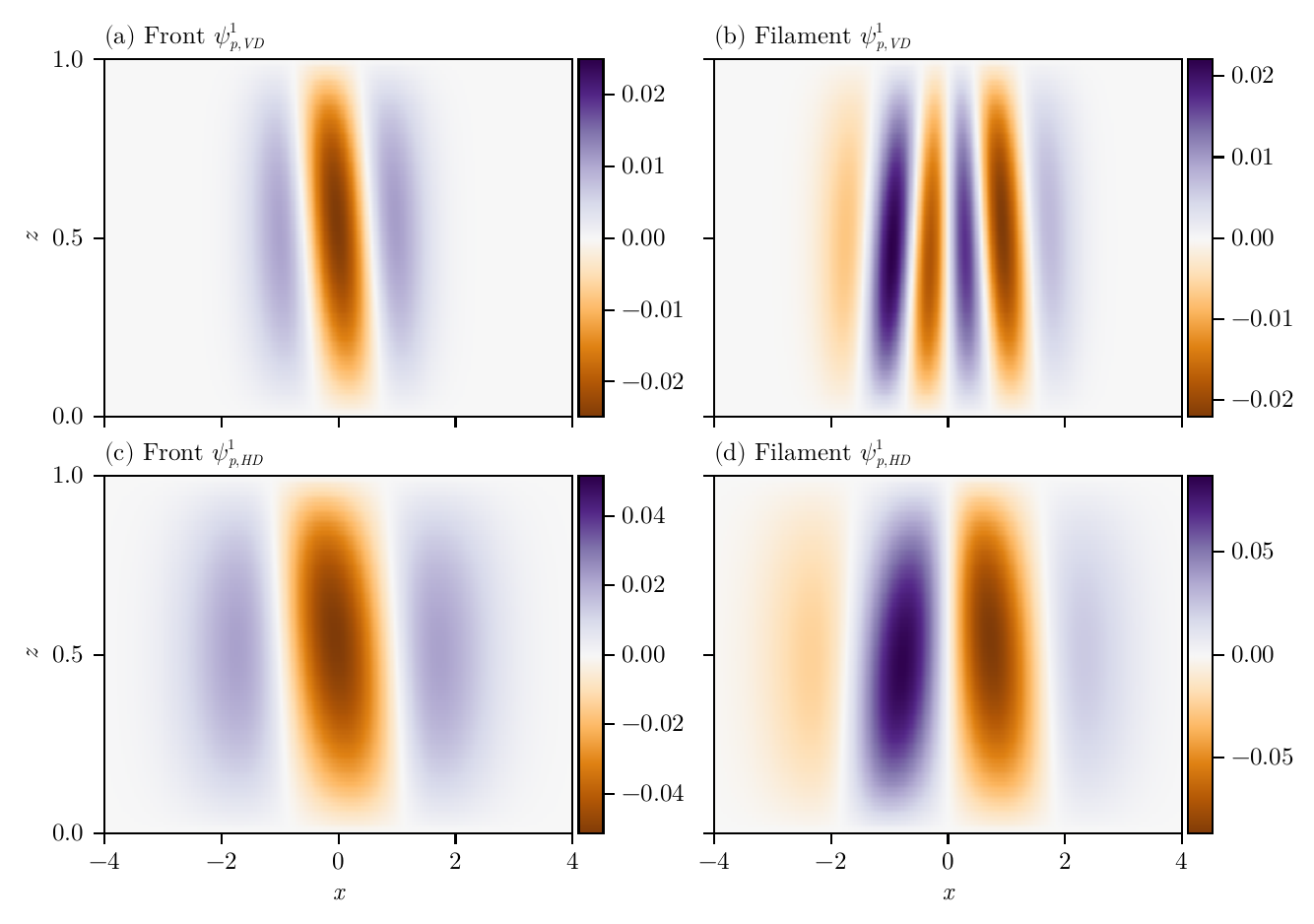}}
  \caption{Particular solution of the diffusivity streamfunction $\psi^1_{p,VD}$, $\psi^1_{p,HD}$ for front and filament. Their viscosity counterparts have the same spatial pattern with just a flip of sign.}
\label{fig:psiVH}
\end{figure}

The detailed expression is shown in Appendix~\ref{app:blt}. Here figure~\ref{fig:psiVH} shows the particular solutions $\psi^1_{p,VD}$ and $\psi^1_{p,HD}$ for the front and filament. The full solution will have the same spatial pattern with just a time-dependent scaling factor $(1-\cos{t})$. Both have a negative cell at the front center that tends to destratify the fluid, while two weaker positive cells are present on the side. The vertical component $\psi^1_{p,VD}$ looks narrower than its horizontal counterpart $\psi^1_{p,HD}$, and the filament case resembles those of two fronts combined with opposite signs. This is consistent with the fact that buoyancy gradients of the two cases are almost horizontal (Figure~\ref{fig:gb1}), so that their vertical components are more constrained in narrow bands. The horizontal diffusivity case (figure~\ref{fig:psiVH}c) looks qualitatively similar to that in B19. Yet note that the forcing here is purely BLT that acts through an ageostrophic residual between the buoyancy and along-front flow, that is, $-\p v^1/\p z+\p b^1/\p x\neq 0$. In terms of streamfunction forcings due to vertical and horizontal viscosities, they will have the same expression with signs flipped from those of diffusivities based on \eqref{eq:fvhdv}. 

\begin{figure}[tp]
\centerline{\includegraphics[width=0.88\linewidth]{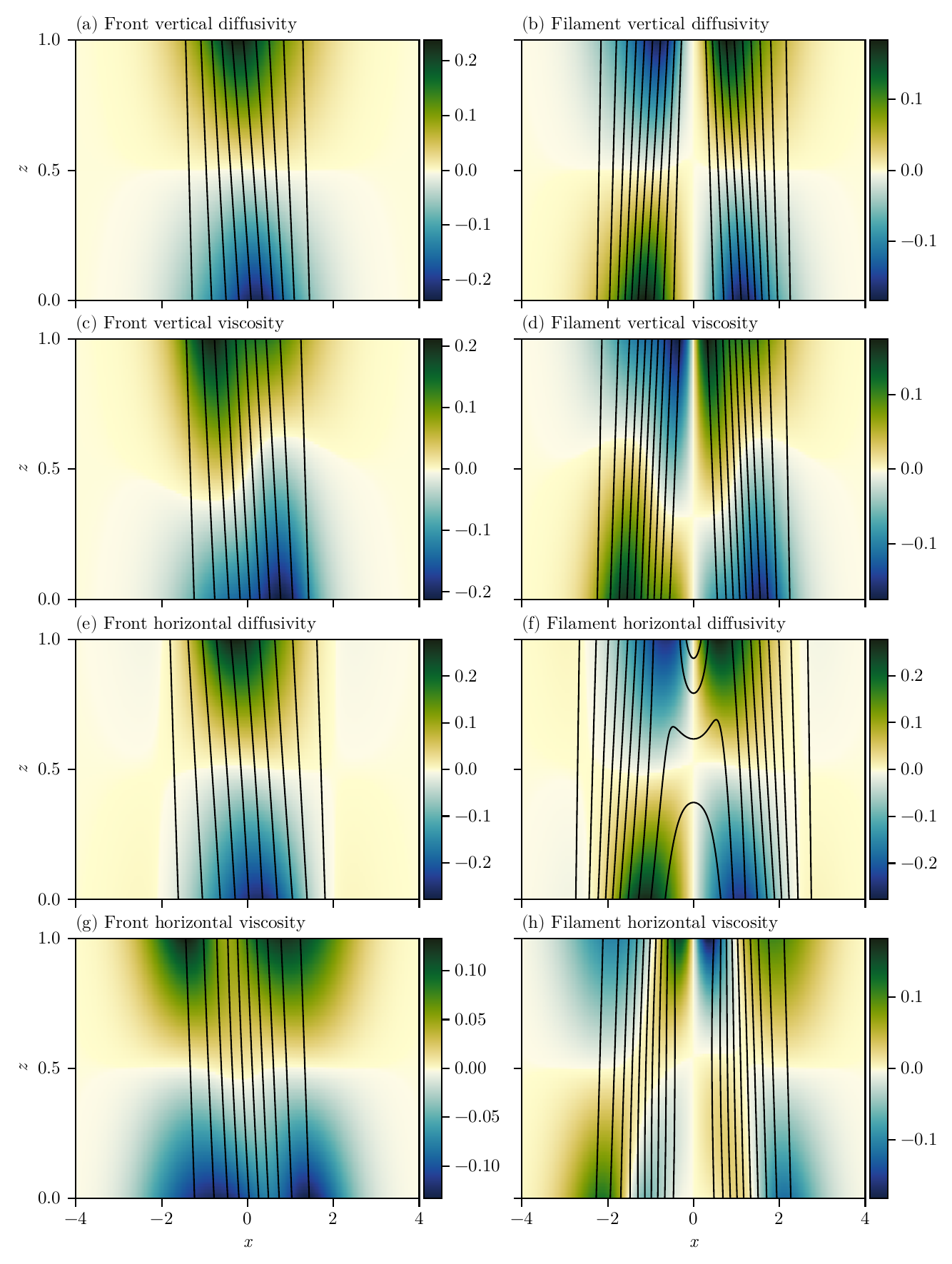}}
  \caption{Front and filament profiles of velocity solutions $v^0+\varepsilon_i v^1_i$ with buoyancy contours $b^0+\varepsilon_i b^1_i$ for (a,b) vertical diffusivity $i=VD$, (c,d) vertical viscosity $i=VV$, (e,f) horizontal diffusivity $i=HD$, (g,h) horizontal viscosity $i=HV$. The relevant parameter is set to $\varepsilon_i =0.03$ with $\Ro=1$, and the time corresponds to $t=10$.}
\label{fig:vVH}
\end{figure}

We can obtain general approximate solutions again using $\psi^1 \approx (1-\cos{t})\psi^1_p$. We confirm the respective errors are small in Appendix~\ref{app:blt}, although they are larger than in the strain-only case. Note that we do not expect a TTW-like circulation since the turbulence forcing gives rise to a linear growth of $v^1$ and $b^1$ to counteract the zeroth-order solution, while the streamfunction feedback is simply oscillatory. In particular, we summarize all relevant terms of boundary layer turbulence effects on velocity and buoyancy in Appendix~\ref{app:blt} and illustrate them in figure~\ref{fig:blt}. 

\begin{figure}[tp]
\centerline{\includegraphics[width=0.88\linewidth]{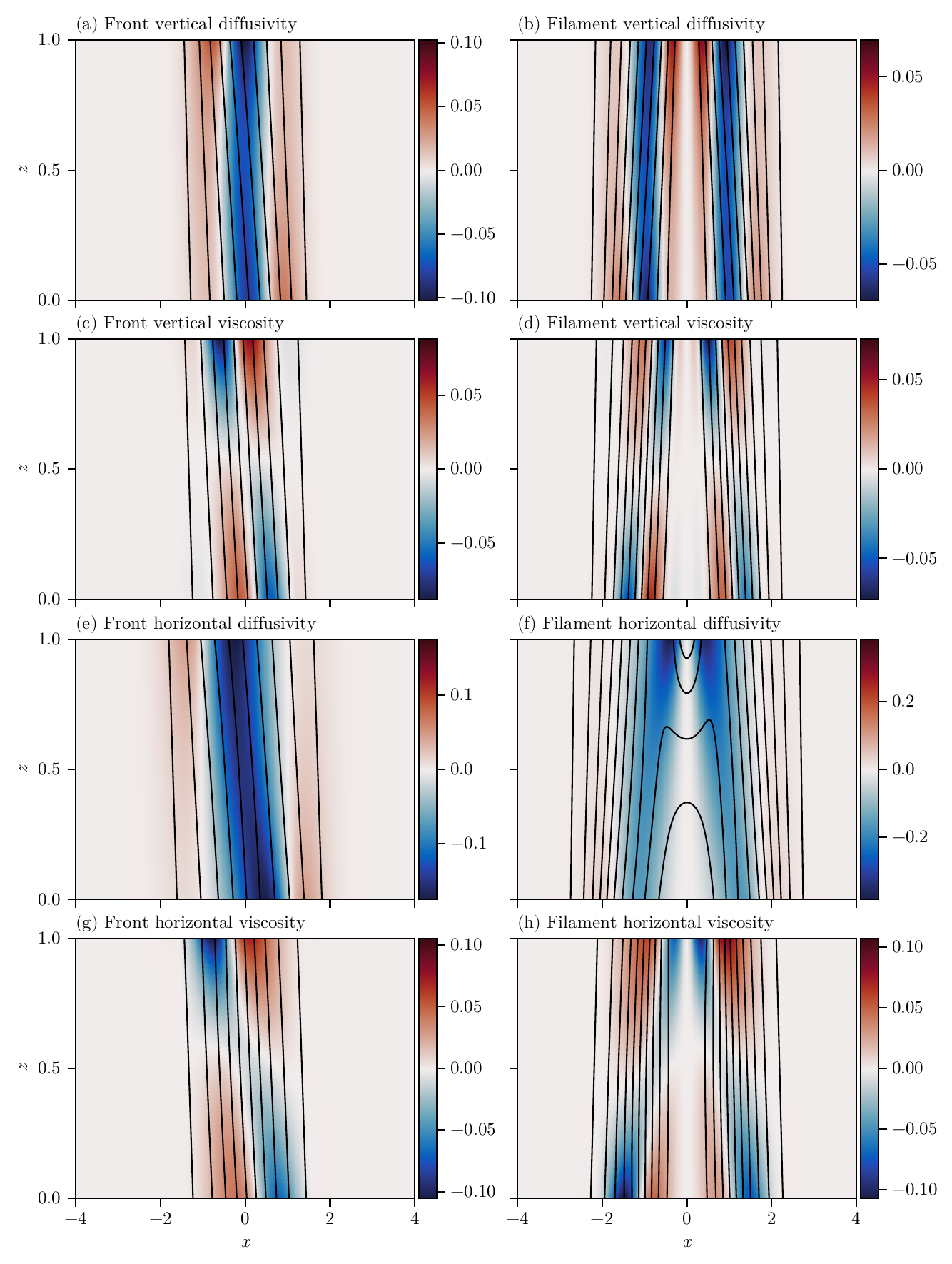}}
  \caption{As in figure~\ref{fig:vVH}, but showing profiles of the first-order frontal tendency $T^1_{b,i}$.}
\label{fig:Tb1VH}
\end{figure}

To isolate the respective effects of horizontal and vertical diffusivity and viscosity, we fix the corresponding forcing parameters to $\varepsilon_i=0.03$, an average of a realistic estimate in Appendix~\ref{app:scaleexp}, with $i\in\{HV,VV,HD,VD\}$ while keeping all others as zero. We discuss a larger parameter exploration in the following section and compare it with strain. Figures~\ref{fig:vVH} and \ref{fig:Tb1VH} illustrate the resulting velocity, buoyancy, and frontal tendency structures at $t=10$.

The vertical diffusivity has a negligible impact on the along-front velocity structure for both the front and filament cases. However, the frontal tendency reveals weak frontolysis in the interior and frontogenesis near the outer isopycnals, structurally similar to the strain-driven cases (figures~\ref{fig:vVH}a,b and \ref{fig:Tb1VH}a,b). In contrast, vertical viscosity introduces significant vertical shear, skewing the velocity distribution such that the maximum shifts toward the frontal edges and spreads vertically. This effect is particularly pronounced in the filament case, where the center near-surface velocity is strengthened (figure~\ref{fig:vVH}c,d). The corresponding frontal tendency exhibits an asymmetric dipole pattern across isopycnals, which is mirrored across $x=0$ for the filament (figure~\ref{fig:Tb1VH}c,d).

In the horizontal diffusivity regime, the velocity profile remains largely robust in the frontal case, whereas the buoyancy contours exhibit significant broadening (figure~\ref{fig:vVH}e). This diffusive widening is most evident in the filament case, where the central isopycnals are heavily distorted near $x=0$, indicating effective erosion of the filament buoyancy gradient (figure~\ref{fig:vVH}f). This is confirmed by the frontal tendency analysis, which shows the strongest and spatially widest frontolytic signature among all cases (figure~\ref{fig:Tb1VH}e,f). For horizontal viscosity, the velocity structure undergoes a bimodal splitting along the horizontal boundaries (figure~\ref{fig:vVH}g), evolving into a quad-modal structure near the surface in the filament case (figure~\ref{fig:Tb1VH}h). The associated frontal tendency resembles that of the vertical viscosity case but is spatially broader, characterized by enhanced frontogenesis at the surface and frontolysis at depth (figure~\ref{fig:Tb1VH}g,h). In this case, the first-order solution is likely being comparable to the zeroth-order solution, such that the perturbation assumption become less accurate at $t=10$.

\begin{figure}[tp]
  \centerline{\includegraphics[width=\linewidth]{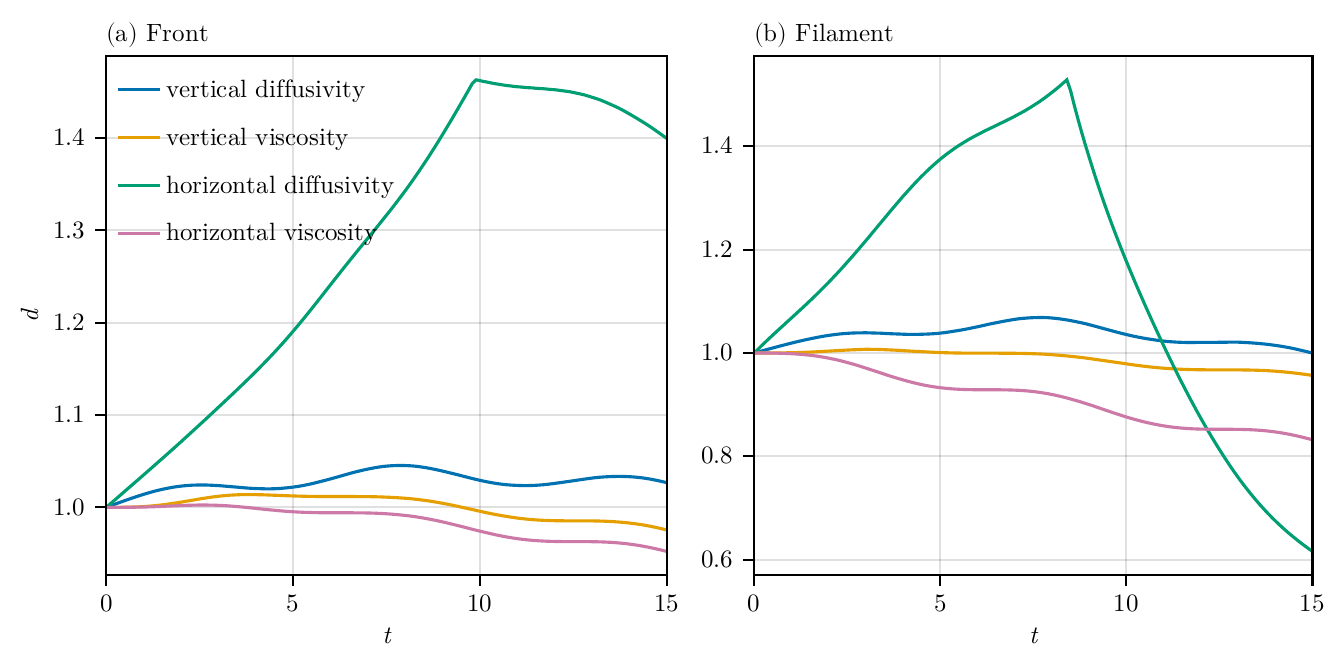}}
  \caption{Evolution of the front width under various turbulent forcing terms for (a) front and (b) filament. The relevant parameter is set to $\varepsilon_i =0.03$ with $i\in\{HV,VV,HD,VD\}$ and $\Ro=1$.}
\label{fig:tsb_ff}
\end{figure}

Figure~\ref{fig:tsb_ff} illustrates the temporal evolution of the frontal width under these distinct turbulent regimes. In the case of vertical diffusivity, the width remains quasi-steady, exhibiting minor oscillations consistent with the balance between competing frontogenetic and frontolytic tendencies. Conversely, vertical viscosity induces a transient widening followed by a gradual contraction, a behavior likely governed by the complex spatial interplay of the tendency terms. The most notable evolution occurs under horizontal diffusivity: the front initially widens significantly, peaking at $t \approx 10$ for the front and slightly earlier for the filament, before undergoing a rapid contraction. This reversal is particularly pronounced in the filament case, where the width collapses to its minimum value in all cases by $t=15$ (figure~\ref{fig:tsb_ff}b). This non-monotonic behavior suggests that while diffusion initially spreads the front, the eventual dominance of frontolysis promotes the emergence of new internal buoyancy gradients, effectively re-sharpening the feature and likely making the perturbation limit less accurate. Finally, horizontal viscosity leads to a monotonic reduction in width over time, aligning with the broadly expanded frontogenetic pattern previously identified. The divergence in these evolutionary pathways suggests that the compound effects of boundary layer turbulence yield a highly dynamic and complex frontal response.


Previous studies have established that vertical mixing, comprising both diffusivity and viscosity, can either induce or inhibit submesoscale frontogenesis \citep{dauhajre_vertical_2025}. We investigate this phenomenon using our first-order model under constant coefficients, yet we do not observe a similar transition (figure~\ref{fig:ro2ekV_sff}a). Therefore, in the asymptotic setting, the transition is likely sensitive to spatial variabilities of turbulence coefficients or higher-order effects \citep{dauhajre_vertical_2025}. 

\subsection{Competition between strain and boundary-layer turbulence} 

\begin{figure}[htpt]
  \centerline{\includegraphics[width=\linewidth]{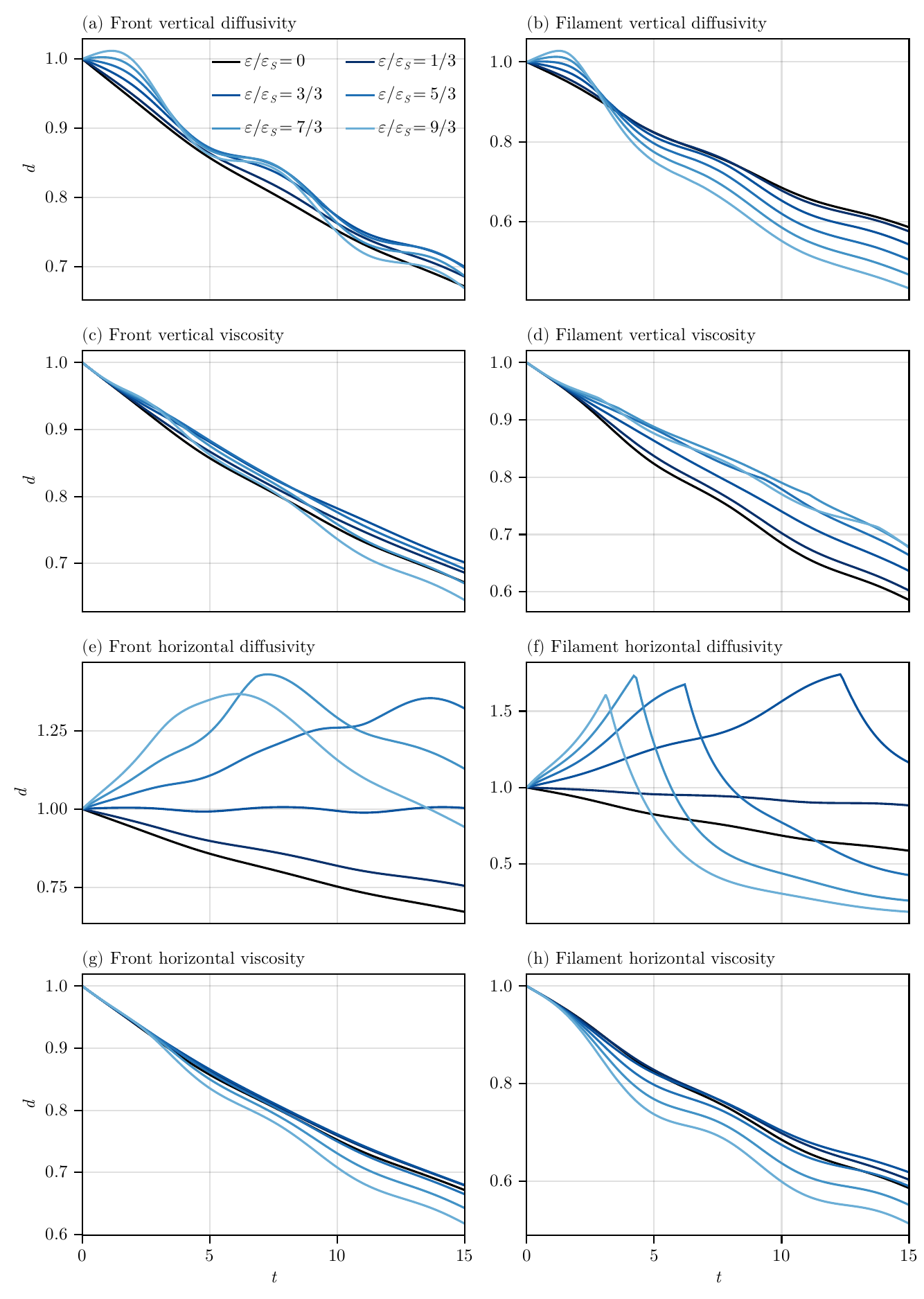}}
    \caption{Evolution of strain-induced frontogenesis under various turbulent forcing terms for front (left column) and filament (right column) with $\Ro=1$. Shown are time series with (a,b) vertical diffusivity, (c,d) vertical viscosity, (e,f) horizontal diffusivity, and (g,h) horizontal viscosity. The strain-only case is in black, and dark to light blues denote an increase of the turbulent scale $\varepsilon$ as $1/3\varepsilon_S, \varepsilon_S, 5/3\varepsilon_S, 7/3\varepsilon_S, 3\varepsilon_S$ with $\varepsilon_S=0.03$.}
\label{fig:tsb4_ff}
\end{figure}

Since perturbations arising from mesoscale strain and boundary-layer turbulence (BLT) frequently coexist, we investigate the frontal response to their concurrent forcing. We initially fix the strain parameter at $\gamma=\varepsilon_S=0.03$, equal to the average BLT parameter used previously, and vary the magnitude of the BLT fluxes (figure~\ref{fig:tsb4_ff}). Later we also present results for a wider range of strain values. For each BLT component, the respective perturbation parameter $\varepsilon$ is increased from 0 up to $3\varepsilon_S$. Horizontal diffusivity exerts the strongest modulation on the frontogenetic process, whereas other fluxes serve to either enhance or mitigate frontal sharpening depending on the temporal phase and the specific frontal configuration.

Specifically, vertical diffusivity introduces inertial oscillations to the strain-induced frontogenesis. In the frontal configuration, this results in a marginally reduced rate of sharpening (figure~\ref{fig:tsb4_ff}a). Conversely, for the filament, vertical diffusivity consistently enhances sharpening for $t > 3$, despite an initial period of broadening similar to the frontal case (figure~\ref{fig:tsb4_ff}b). This likely stems from the structural alignment between the frontal tendency patterns induced by strain and those induced by vertical diffusivity in the filament geometry (cf. figure~\ref{fig:Tb1VH}b). Vertical viscosity, however, consistently inhibits filament sharpening. For the front, moderate viscosity scales lead to widening, while the largest magnitude ($\varepsilon = 3\varepsilon_S$) induces sharpening after $t \approx 10$. This suggests that the bimodal surface tendency patterns associated with vertical viscosity generally counteract strain-driven frontogenesis, although extreme viscous forcing can eventually dominate the dynamics to favor frontogenesis (figure~\ref{fig:Tb1VH}c,d).

Horizontal diffusivity is unique in its ability to trigger transient frontolysis even in the presence of strain (figure~\ref{fig:tsb4_ff}e,f). For the front, a diffusivity scale of $\varepsilon = \varepsilon_S/3$ significantly weakens sharpening, while $\varepsilon = \varepsilon_S$ yields a quasi-steady width, and stronger diffusion ($\varepsilon > \varepsilon_S$) leads to broadening over finite times (figure~\ref{fig:tsb4_ff}e). The filament case exhibits similar but more pronounced behavior; notably, at later times, strong horizontal diffusivity can trigger sharper widths than the strain-only reference case (figure~\ref{fig:tsb4_ff}f). Finally, horizontal viscosity demonstrates a weak mitigating effect at small magnitudes but becomes dominantly frontogenetic at large magnitudes, a trend that is consistent across both front and filament configurations (figure~\ref{fig:tsb4_ff}g,h).

\begin{figure}[htbp]
  \centerline{\includegraphics[width=\linewidth]{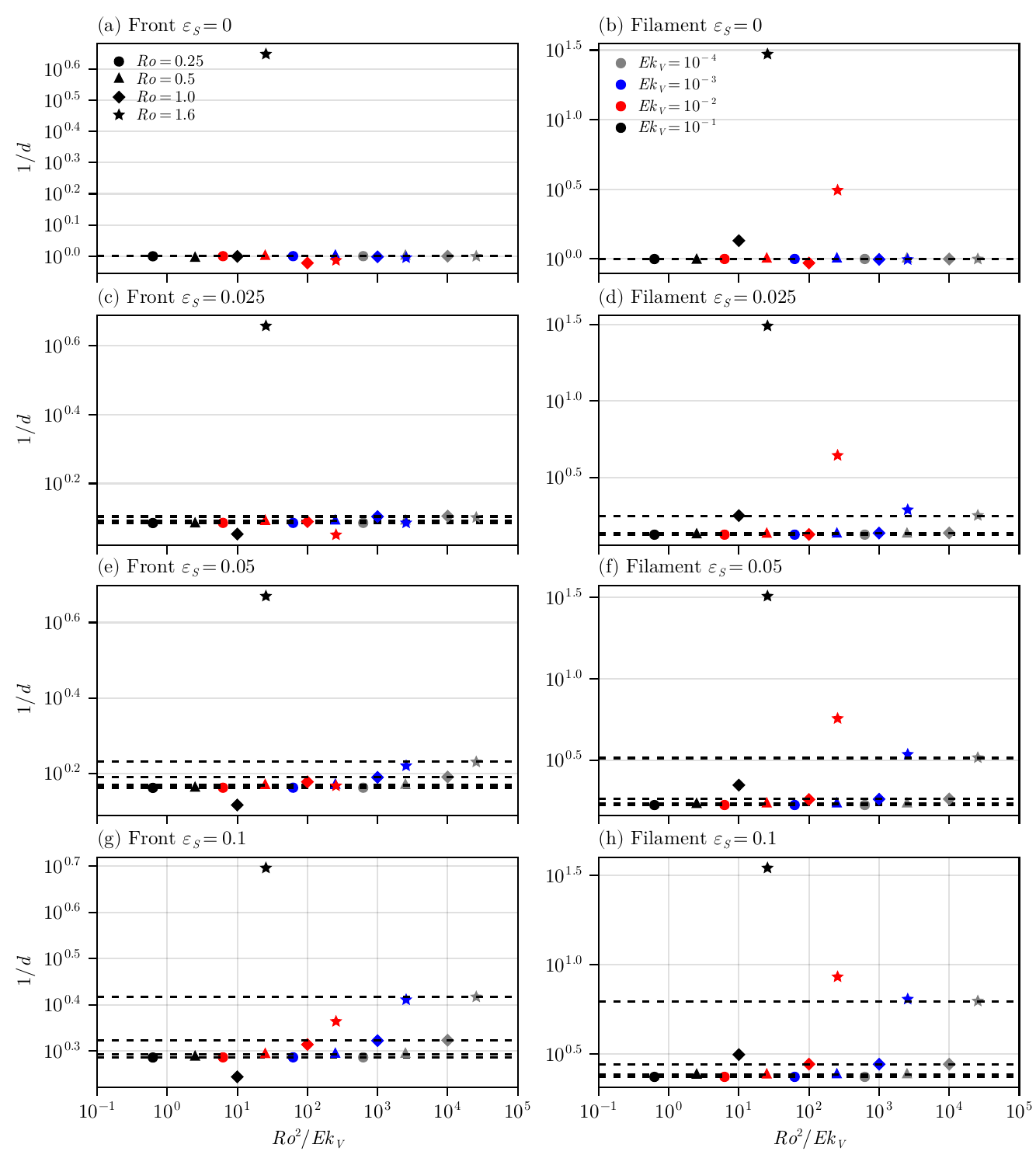}}
  \caption{$d$-based metrics of frontogenesis $1/d$ as a function of $\Ro^2/\Ek_V$ under varying strain intensities. Shown are (a,b) $\varepsilon_S=0$, (c,d) $\varepsilon_S=0.025$, (e,f) $\varepsilon_S=0.05$, and (g,h) $\varepsilon_S=0.1$. The left column corresponds to the front, and the right column to the filament. Horizontal dashed lines indicate the reference frontogenesis level for $\Ek_V=0$ at the corresponding $\Ro$; the first row ($\varepsilon_S=0$) serves as a baseline where this reference is zero. Here we ignore horizontal mixing with $\Ek_H=0$ throughout.}
\label{fig:ro2ekV_sff}
\end{figure}

\begin{figure}[htbp]
  \centerline{\includegraphics[width=\linewidth]{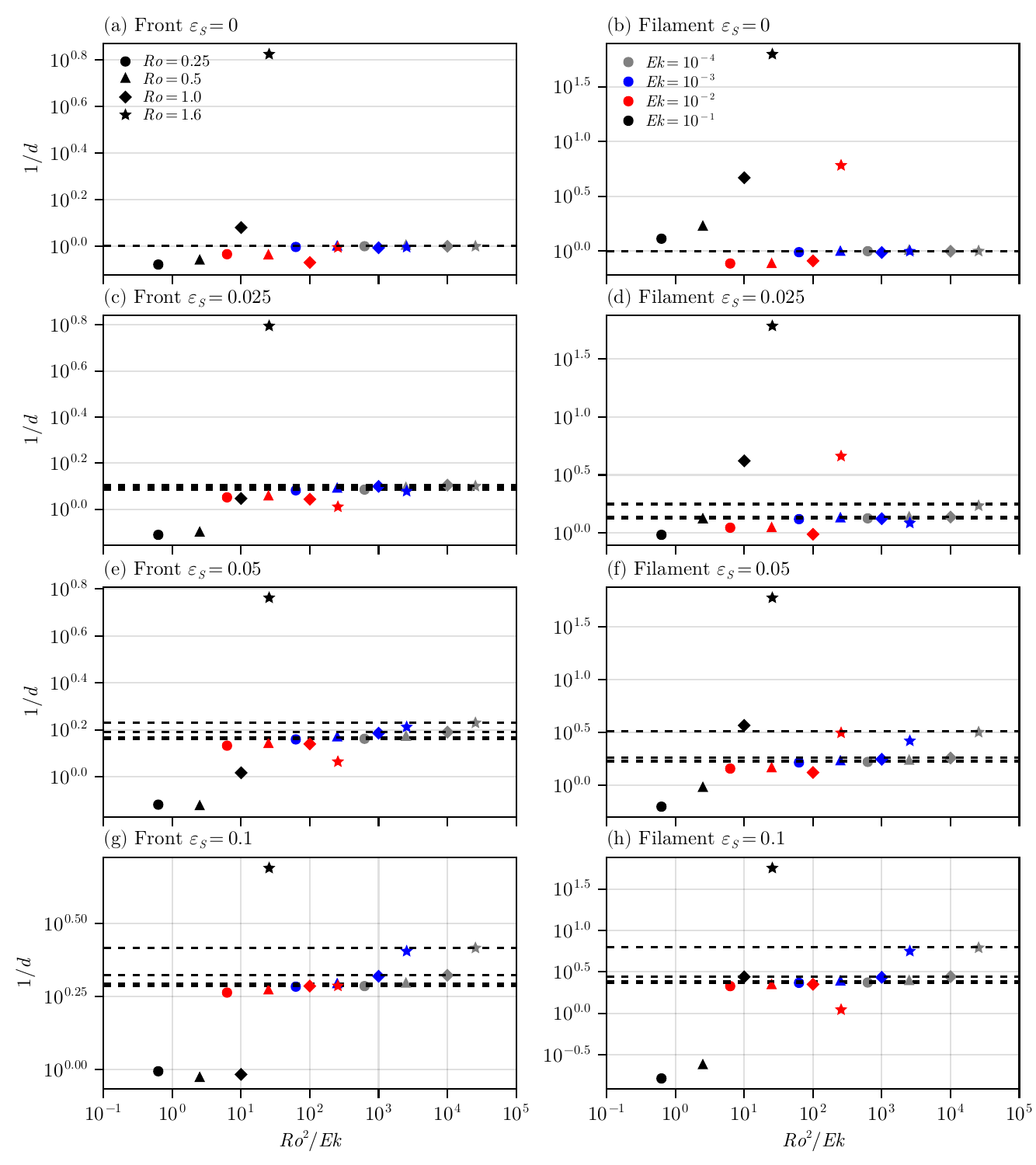}}
  \caption{Same as figure~\ref{fig:ro2ekV_sff} but with $\Ek=\Ek_V=\Ek_H$ and $\Pran_V=\Pran_H=1$.}
\label{fig:ro2ek_sff}
\end{figure}

Building on recent work by \citet{dauhajre_vertical_2025}, we further quantify the combined impact of vertical diffusivity and viscosity under varying strain conditions across the $\Ro$--$\Ek_V$ parameter space with $\Pran_V=1$, and $\Ek_H=0$ (figure~\ref{fig:ro2ekV_sff}). The parameter space explored is $\Ro \in \{0.25, 0.5, 1.0, 1.6\}$ and $\Ek_V \in \{10^{-4}, 10^{-3}, 10^{-2}, 10^{-1}\}$, mirroring the regime studied by \citet{dauhajre_vertical_2025}, with the exception of $\Ro=1.6$, which is smaller than their choice of 2 to avoid super-frontogenesis under strain. We calculate the frontogenesis metric $1/d$ at $t=10$ (approximately 1.5 inertial periods), following \citet{dauhajre_vertical_2025}. 

In the absence of strain ($\varepsilon_S=0$), the $1/d$ values remain largely at unity across the parameter space for both configurations, corresponding to lack of frontogenesis (figure~\ref{fig:ro2ekV_sff}a,b). Exceptions appear at $(\Ro,\Ek_V)=(1.6,10^{-1})$ for front and $(1.6,10^{-1}),(1.6,10^{-2}),(1.0,10^{-1})$ for filament, which shows frontogenetic levels significantly higher than unity. Under weak strain ($\varepsilon_S=0.025$), the reference inviscid cases ($\Ek_V=0$) exhibit comparably low levels of frontogenesis across all Rossby numbers (figure~\ref{fig:ro2ekV_sff}c,d). The inclusion of vertical mixing generally results in values overlapping with these background levels, excepting super-frontogenetic outliers $(\Ro,\Ek_V)$ combinations same as those in the zero-strain case. However, the tendency is significantly weakened for the front scenario at $(\Ro, \Ek_V)=(1.0,10^{-1}),(1.6,10^{-2})$. 

As the strain increases to moderate levels ($\varepsilon_S=0.05$), the inviscid reference levels become distinctly stratified by Rossby number, with higher $\Ro$ yielding stronger frontogenesis (figure~\ref{fig:ro2ekV_sff}e,f). The previously identified deviations persist but with altered relative magnitudes; for instance, the front at $\Ro=1.0, \Ek_V=10^{-1}$ remains the most inhibited case, while cases at $\Ek_V=10^{-2}$ show similar frontogenetic levels. Under strong strain ($\varepsilon_S=0.1$), the stratification of the inviscid reference levels is further amplified (figure~\ref{fig:ro2ekV_sff}g,h). In this regime, higher $\Ro$ gives stronger frontogenesis for frontal cases at $\Ek_V=10^{-2}$, although its values at $\Ro=1.6$ remains substantially weakened than the baseline. That said, a step-like pattern start to emerge in the frontal case, while other patterns remain similar to those under moderate strain. If we compare the results to the viscosity-only case with $\Pran_V=\infty$, as explored in \citet{dauhajre_vertical_2025}, the frontogenetic strength are generally stronger when diffusivity is present, except at $\Ro=1.0, \Ek_V=10^{-1}$ (figure~\ref{fig:ro2ekV_vsff}). Therefore, in this framework the weak-strong frontogenesis division is governed by the $\Ro$-sensitive strain effect, in contrast to the diffusivity-viscosity competition discussed in \citet{dauhajre_vertical_2025} which neglects strain.

These results underscore the complex, non-monotonic interplay between strain-driven frontogenesis and vertical mixing, revealing a high sensitivity to the specific dynamical regime defined by the Rossby and Ekman numbers. And based on the analysis under a fixed strain and varied degrees of each BLT flux (figure~\ref{fig:tsb4_ff}), we expect more sophisticated interactions once horizontal mixing is also considered. Here we demonstrate this by setting $\Ek=\Ek_V=\Ek_H$ with $\Pran_V=\Pran_H=1$ and repeat the $\Ro$--$\Ek$ parameter space exploration as done in figure~\ref{fig:ro2ekV_sff}, so all terms of horizontal and vertical viscosity and diffusivity are present at equal magnitudes. As expected, the results show a more robust step-like transition for both front and filament, and the frontogenetic level $1/d$ can drop below unity. However, if we consider only viscosities by setting $\Pran_V=\Pran_H=\infty$, the frontolytic cases with $1/d$ significantly below unity disappear (figure~\ref{fig:ro2ek_vsff}). Therefore, in this case certain frontolytic-frontogenetic dynamic partitions similar to that in \citet{dauhajre_vertical_2025} are emerging (figure~\ref{fig:ro2ek_sff}), yet the difference of adding horizontal mixing illustrate the effect of $\Ek_H$, and the difference of removing diffusivities implies that horizontal diffusivity dominates the frontolytic evolution than any other terms. Therefore, horizontal mixing is critical in this perturbation model for frontolytic evolution, as suggested already in B19.


\section{Discussion and conclusions}
\label{sec:discussion}

In this study, we have performed a perturbation analysis of initially geostrophically balanced 2D fronts and filaments subject to comparable mesoscale strain and boundary layer turbulence (BLT), a regime typical of weakly stratified oceanic mixed layers. The initial configuration adapts the analytical solution of \citet{shakespeare_generalized_2013}, assuming zero PV and a finite Rossby number typical of submesoscale dynamics. This approach complements the semi-geostrophic analysis of \citet{bodner_perturbation_2019} (hereafter B19), enabling a systematic examination of frontal characteristics across a broad range of Rossby, Ekman, and normalized strain numbers. We note that while some submesoscale fronts can extend below the mixed layer \citep{siegelman_enhanced_2020}, our analysis is confined to the mixed-layer alone.

When both strain and BLT are treated as perturbations, we assume the zeroth-order solution as a geostrophic balance that permits finite Rossby numbers. The first-order perturbation solution is substantially simplified by initializing the zeroth-order state in geostrophic balance. This allows the system to be reduced to one involving only the first-order cross-front streamfunction, which we solve using momentum-following coordinates and normal-mode analysis. For pure strain perturbations, our solution is closely related to that of \citet{shakespeare_generalized_2013}; indeed, treating the strain parameter as a perturbation in their analysis would yield a matching result, albeit with a non-exponential but persistent frontal tendency. In the early inertial period, these approximations show strong agreement. In the real ocean, the effect of BLT can be comparable to and limit strain-induced frontogenesis, in which case the linear perturbation solution may offer a more physically realizable description than the full exponential growth predicted by inviscid theories.

For BLT perturbations acting on the geostrophic base state, the response is diverse across horizontal versus vertical components and dissipation versus viscosity. Consistent with B19, we find that the horizontal component is predominantly frontolytic, whereas vertical viscosity drives frontogenesis. However, in contrast to \citet{dauhajre_vertical_2025}, who identified a transition between frontolytic and frontogenetic regimes for vertical mixing depending on the Rossby--Ekman number combination, our perturbation analysis, assuming uniform viscosity and diffusivity, reveals that all relevant scenarios are frontogenetic. This difference highlights the sensitivity of the solution to the specific formulation of the turbulent fluxes and the base state. To maintain analytical tractability, we employ constant eddy diffusivity and viscosity, yet the formulation is robust enough to accommodate spatially inhomogeneous turbulent boundary layers in future extensions.

When finite-Rossby-number fronts and filaments are perturbed simultaneously by strain and BLT, vertical viscosity and diffusivity can attenuate strain-induced frontogenesis in certain parameter regimes, despite the inherent frontogenetic tendency of vertical viscosity. Critically, horizontal diffusivity acts as a potent stopper of frontogenesis and can counterbalance strain-induced sharpening when perturbation scales are comparable. Conversely, horizontal viscosity serves to accelerate frontal sharpening. Exploring the parameter space of Rossby, Ekman, and normalized strain numbers reveals outlier cases where BLT significantly modulates the strain impact. Specifically, in the vertical mixing scenario at the highest Rossby and Ekman numbers, turbulence generally enhances frontogenesis. However, at slightly lower but still order-unity Rossby numbers, this effect reverses, weakening the strain impact. At moderate Rossby and Ekman numbers, the interaction is variable, alternating between weakening and strengthening. This complex, non-monotonic interaction highlights the necessity of carefully characterizing both strain and BLT when parameterizing submesoscale dynamics \citep{zhang_2022_parameterizing, feng_implementation_2026}.

Our results also show broad consistency with recent Large Eddy Simulations (LES). For instance, \citet{sullivan_oceanic_2024} observed that boundary layer turbulence could arrest frontogenesis, a feature we capture analytically through the action of horizontal diffusivity. Furthermore, our asymptotic approach differs from and is complementary to other methods such as diagnostic force-balance decompositions or turbulent thermal wind (TTW) analyses \citep{mcwilliams_submesoscale_2017, crowe_evolution_2018}. Diagnostic approaches, by their nature, do not capture the temporal evolution, while previous TTW analyses typically assume a small Rossby number. Our asymptotic framework is designed to illuminate different facets of the frontal dynamics. Our latest LES work explore comprehensive energetic interactions among submesoscale and BLT under spatially heterogeneous mesoscale forcing \citep{peng_submesoscale_2026}. The mesoscale field holds a range of strain effects, from strong convergence to divergence. The novelty of our approach here lies in decoupling these effects analytically to reveal their individual and competitive contributions across a wide parameter space. Therefore, in principle in can be applied to cases under negative strain (divergence), which is less explored but not uncommon in upper ocean multiscale dynamics.

The parameterized turbulent fluxes in our framework can, in principle, represent mixed-layer instability (MLI), which restratifies the mixed layer by releasing available potential energy from horizontal buoyancy gradients \citep{boccaletti_mixed_2007}. Recent analyses suggest that MLI exhibits characteristics of both baroclinic and barotropic instability \citep{kar_linear_2025}. Our results can be extended to include a more detailed treatment of turbulent fluxes arising from both surface forcing (BLT) and submesoscale instabilities. This has significant implications for the development of coupled BLT-submesoscale parameterizations in ocean models.

The intricate evolution of the frontal tendency under combined strain and BLT forcing suggests that existing, simplified parameterization schemes may fail to capture key interactions in realistic, ageostrophic fronts and filaments. In particular, the spatial sensitivity of the BLT-induced frontal tendency to specific viscosity and diffusivity profiles is often underrepresented in mixed-layer parameterizations \citep[e.g.,][]{bodner_modifying_2023}. This underscores the critical need for revised parameterization frameworks that incorporate front-aware spatial variability to more accurately represent upper-ocean frontal evolution.

\section*{Declaration of interests}{The authors report no conflict of interest.}
%
\section*{Data availability statement}{The code that supports the findings of this study is openly available in \url{https://github.com/bodner-research-group/LESStudySetup.jl}. In accordance with the journal's policy, the authors declare the use of generative AI tools, specifically ChatGPT and Gemini, to assist with code debugging and the linguistic refinement of the text originally drafted by the authors. The authors have reviewed all AI-generated output and assume full responsibility for the accuracy and integrity of the final manuscript.}
%
%

\begin{appen}
\section{Perturbation scaling demonstration example}\label{app:scaleexp}
 Using typical submesoscale parameters of $f=\qty{1e-4}{\per\second}$, $L=\qty{1}{\kilo\meter}$, $\Ro=1$ indicates a reasonable velocity scale of $\qty{0.1}{\meter\per\second}$ in the upper ocean. Assuming a typical Ekman depth $d=\sqrt{2\bar \nu_V/f}=\qty{10}{\meter}$, we have $\bar \nu_V=\qty{5e-3}{\meter\squared\per\second}$. This is much greater than the molecular viscosity of water, so we expect the viscosities to reasonably involve parameterized turbulence. By setting $H=\qty{50}{\meter}$ and $\Ek_H=\Ek_V$, we get $\varepsilon_{HV}=\varepsilon_{VV}=0.02$, $\bar \nu_H=\qty{2}{\meter\squared\per\second}$, and $M^2=\qty{2e-7}{\per\second\squared}$. With a typical mesoscale strain magnitude
$\alpha=\qty{1e-6}{\per\second}$, we have 
$\varepsilon_S=0.01$. Assuming $\Pran_V=\Pran_H=0.5$, 
we have $\varepsilon_{HD}=\varepsilon_{VD}=0.04$, $\bar \kappa_V\approx\qty{0.01}{\meter\squared\per\second}$ and $\bar \kappa_H\approx\qty{2}{\meter\squared\per\second}$. 

\section{First-order strain-induced streamfunction and approximation error}\label{app:strainerror}
This is consistent with the theory in ST13. In particular, if we treat the scaled strain $\gamma$ as the small parameter, and derive a corresponding analytical first-order solution based on the assumptions in ST13 and under the full momentum coordinate $X =  e^{\gamma T}(x+ \Ro^2\, v)$, the general solution of the streamfunction with geostrophically balanced initial state 
\begin{equation}\label{eq:psifull}
    \psi_{ST} =  -\Ro \,{B_0'(X)}Z(Z-1)e^{\gamma T}\left[\gamma \left(e^{\gamma T}-\cos{\sqrt{1-\gamma^2}T}\right)-\gamma^2 \frac{\sin{\sqrt{1-\gamma^2}T}}{\sqrt{1-\gamma^2}}\right].
\end{equation}
can be expanded in $\gamma$ as $\psi_{ST} = \psi^0_{ST}+\gamma \psi^1_{ST}+\textit{O}(\gamma^2)$, giving 
\begin{equation}\label{eq:psi1strain}
    \psi^1_{ST} =  -\Ro \,{B_0'(X_0)}Z(Z-1)(1-\cos{T}).
\end{equation}
Note the same spatial function structure ${B_0'(X_0)}Z(Z-1)$ between \eqref{eq:psi1p} and \eqref{eq:psi1strain}, and the temporal component in \eqref{eq:psi1strain} with inertial frequency that is consistent with the general normal mode analysis. These are only approximate solutions, since the derivation in ST13 ignored the nonlinear vertical advection that involves one first order term in this perturbation analysis. One can verify that these solutions satisfy the evolution equations for $v^1$ and $b^1$ but not $\psi^1$. Similar to ST13 that confirmed in the appendix that the nonlinear terms are small, we also confirm here that the error is higher-order for the evolution of $\psi^1$.

\begin{figure}
\centerline{\includegraphics[width=\linewidth]{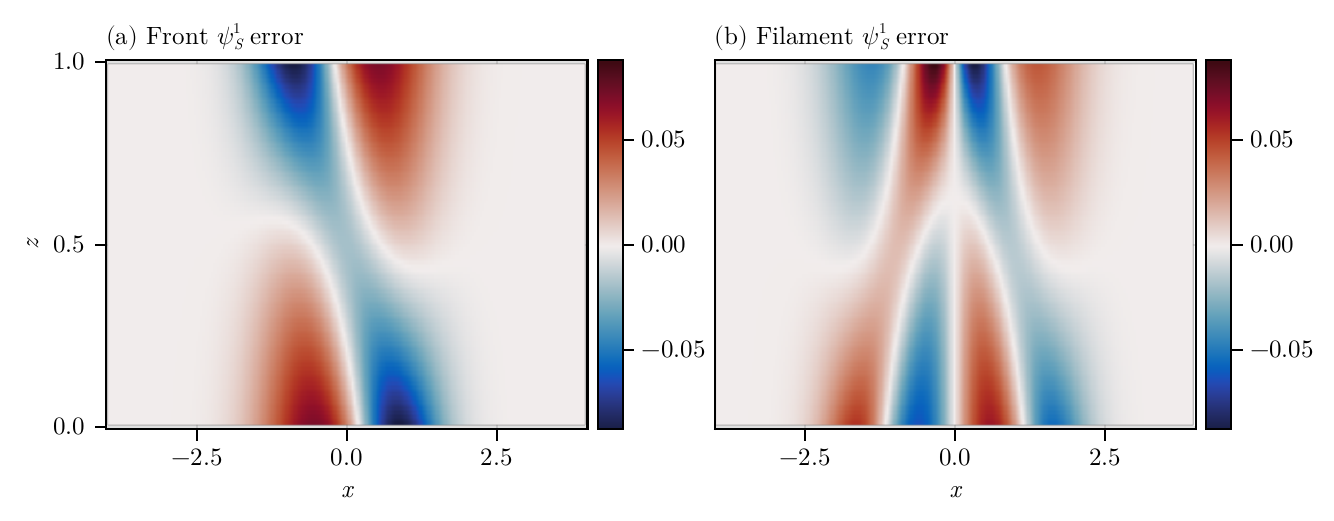}}
  \caption{Relative error of first-order strain-induced streamfunction.}
\label{fig:psig_e}
\end{figure}
Here we assess the error made in using an expansion in ST13 as the first-order solution for the strain-only case. Figure~\ref{fig:psig_e} shows that the relative error is less than $0.1$ in magnitude, and the extrema concentrates on the front edges at the boundaries. Therefore, it is reasonable analytical approximation. 


\begin{figure}
\centerline{\includegraphics[width=\linewidth]{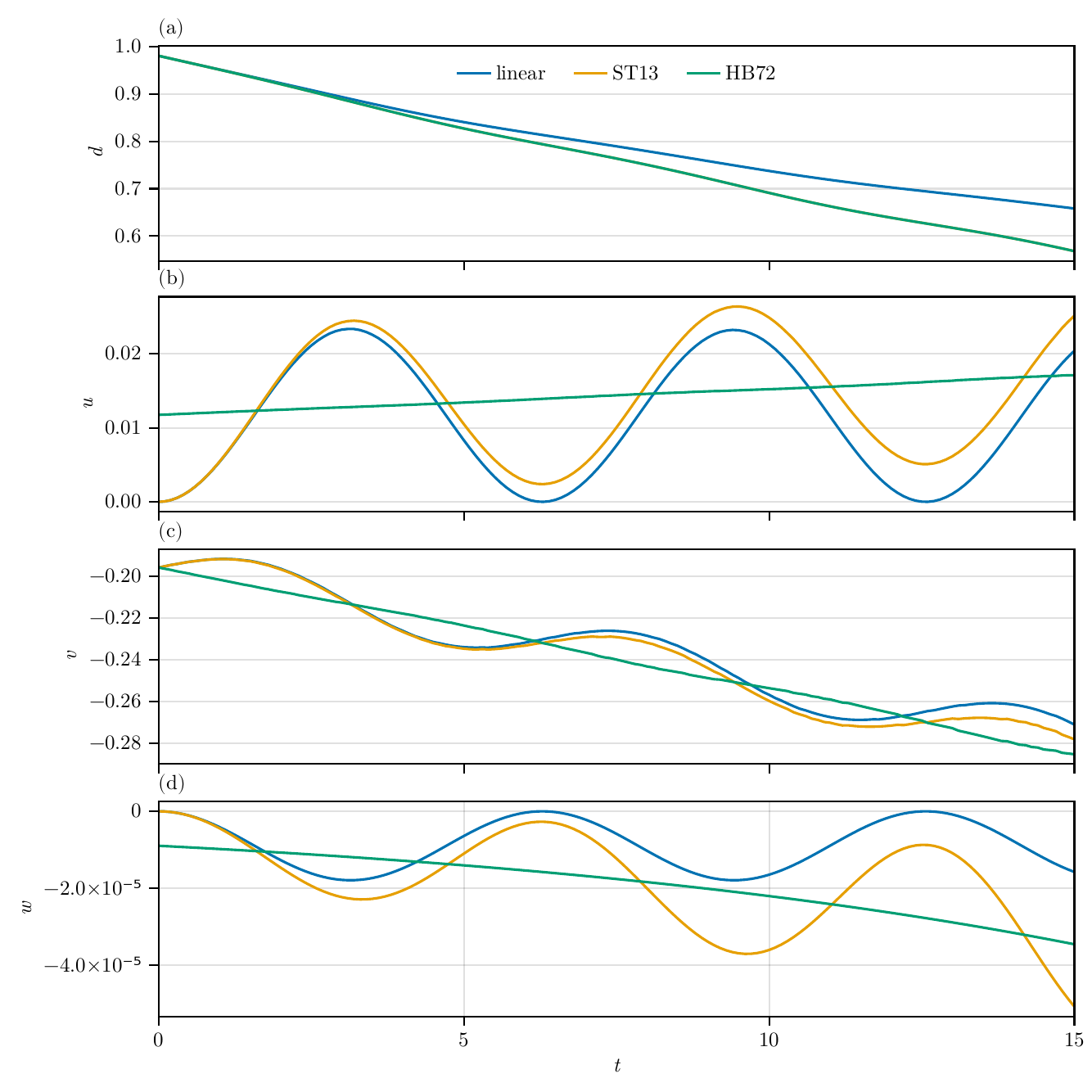}}
  \caption{Time evolution of (a) frontal width $d$, (b) cross-front velocity $u$, (c) along-front velocity $v$, and (d) vertical velocity based on the theory in this study (blue), ST13 (orange), and HB72 (green). They are located at the $x$ position of the buoyancy gradient maximum (the front) for parameter values
  of $\varepsilon_S=0.03$ and $\Ro=1$. The vertical position is $z =0$ for the horizontal velocities $u$ and $v$, and $z =1/2$ for the vertical velocity $w$.}
\label{fig:duvw}
\end{figure}

In terms of time evolution, Figure~\ref{fig:duvw} quantitatively demonstrates the sharpening of the horizontal buoyancy gradient and evolving velocity components as in ST13. The linear solution (zeroth+first-order) proposed here captures the bulk evolution of $d$ and $v$ with less than \qty{10}{\percent} offset for $t<10$. The pure oscillatory nature of $u$ and $w$ in the first-order solution confirms that the exponential feedback from the sharpened front is a higher-order long-term effect. Since we have confirmed that the approximation is a higher-order effect, we focus on the first inertial period ($t<10$), where the first-order solution captures the strain effect fairly well.

\begin{figure}[t]
\centerline{\includegraphics[width=\linewidth]{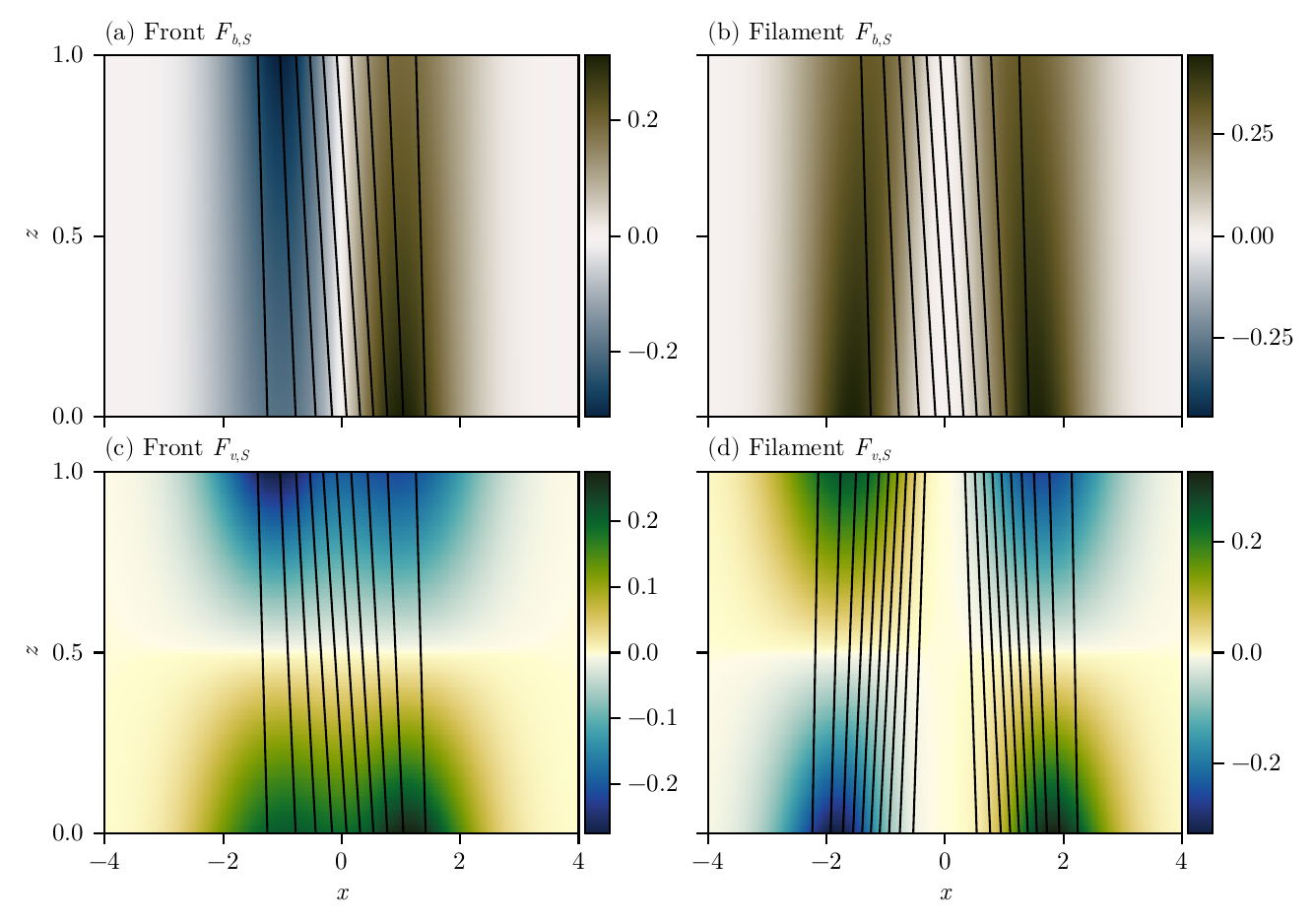}}
  \caption{Buoyancy and velocity forcings under strain. Shown are (a) buoyancy forcing for front, (b) buoyancy forcing for filament, (c) velocity forcing for front, (d) velocity forcing for filament. Parameters used are $\Ro=1$, ${\widetilde \varepsilon_S}=1$ and zero for all BLT terms.}
\label{fig:fbfv}
\end{figure}

\section{Particular solution with boundary layer turbulence}\label{app:blt}
\begin{figure}
  \centerline{\includegraphics[width=\linewidth]{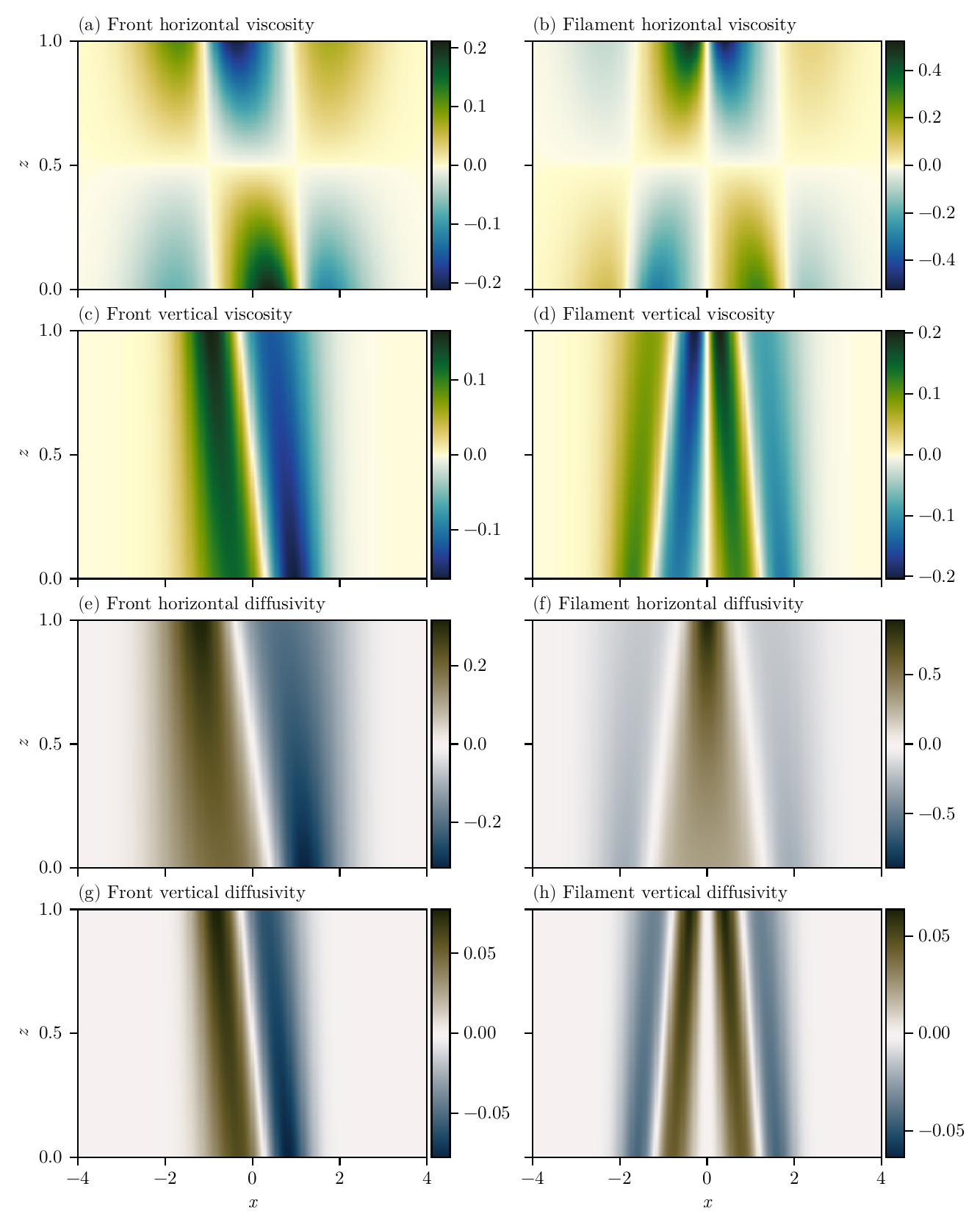}}
  \caption{Boundary layer turbulence terms for front and filament.}
\label{fig:blt}
\end{figure}
Relevant terms in $F_v$ and $F_b$, assuming $\nu_H=\nu_V=\kappa_H=\kappa_V=1$, are
\begin{subequations}
\begin{align}
    \frac{\p^3 \phi^0}{\p x^3}& = J_0^{3}B^{(3)}_0 \left(Z-\frac 12\right),\\
    \frac{\p^3 \phi^0}{\p x\p z^2}& = \Ro^2J_0^{2}B'_0 \left[ 2B''_0 + \Ro^2{B'_0B^{(3)}_0\left(Z-\frac 12\right)}{J_0} \right],\\
    \frac{\p^3 \phi^0}{\p x^2\p z}& =J^2_0\left[ B''_0 + \Ro^2{B'_0B^{(3)}_0}\left(Z-\frac 12\right){J_0} \right],\\
    \frac{\p^3 \phi^0}{\p z^3}& = \Ro^4J_0^{2}(B'_0)^2 \left[ 3B''_0 + \Ro^2{B'_0B^{(3)}_0\left(Z-\frac 12\right)}{J_0} \right].
\end{align}
\end{subequations}
The relevant terms in the streamfunction equation are given by
\begin{subequations}
\begin{align}
    \frac{\p^4 \phi^0}{\p x\p z^3}& = J^4_0\left[A_0(X_0)+A_1(X_0) \left(Z-\frac 12\right)+ A_2(X_0)\left(Z-\frac 12\right)^2\right],\\
    \frac{\p^4 \phi^0}{\p x^3\p z}& =J^4_0\left[B^{(3)}_0(X_0)+A_3(X_0) \left(Z-\frac 12\right)+ A_4(X_0)\left(Z-\frac 12\right)^2\right].
\end{align}
\end{subequations}
where
\begin{subequations}
\begin{align}
    A_0(X_0)& = 3B'_0 \left( 2(B''_0)^2 +B^{(3)}_0 B'_0  \right),\\
    A_1(X_0)& = -\Ro^2\,B'_0 \left[ 12(B''_0)^3 -  B^{(4)}_0 (B'_0)^2-3B^{(3)}_0 B'_0 B''_0 \right],\\
    A_2(X_0)& = \Ro^4\, B'_0 \left[ 6(B''_0)^2\left((B''_0)^2 -B^{(3)}_0 B'_0 \right)- {(B'_0)^2 \left(B^{(4)}_0 B''_0-3(B^{(3)}_0)^2\right)} \right],\\
    A_3(X_0)& = \Ro^2 \left[ B^{(3)}_0B''_0 +  B^{(4)}_0 B'_0 \right],\\
    A_4(X_0)& = -\Ro^4 \left[ 2B^{(3)}_0(B''_0)^2 + {B'_0 \left(B^{(4)}_0 B''_0-3(B^{(3)}_0)^2\right)} \right].
\end{align}
\end{subequations}
The corresponding particular solution is
\begin{equation}\label{eq:psi1p2}
    \psi^1_{V,p}\propto C_1Z+C_2+C_3\Psi(Z)\left(\ln{\frac{\Psi(Z)}{2}}-1\right)+C_4\ln{\Psi(Z)}+ \frac{C_5}{\Psi(Z)},
\end{equation}
where $\Psi(Z) = 2+\Ro^2B'_0 (1-2Z)$ and for $F_V$
\begin{subequations}
\begin{align}
    C_3(X_0)& = \frac{A_2}{2(\Ro^2 B'_0)^4},\\
    C_4(X_0)& = \frac{A_1\Ro^2 B'_0+2A_2}{(\Ro^2 B'_0)^4},\\
    C_5(X_0)& = \frac{A_0(\Ro^2 B'_0)^2+A_1\Ro^2 B'_0+A_2}{(\Ro^2 B'_0)^4},\\
    C_2(X_0)& = -\left[C_3\Psi(0)\left(\ln{\frac{\Psi(0)}{2}}-1\right)+C_4\ln{\Psi(0)}+ \frac{C_5}{\Psi(0)}\right],\\
    C_1(X_0)& = -\left[C_2+C_3\Psi(1)\left(\ln{\frac{\Psi(1)}{2}}-1\right)+C_4\ln{\Psi(1)}+ \frac{C_5}{\Psi(1)}\right],\\
\end{align}
\end{subequations}
\begin{figure}
  \centerline{\includegraphics[width=\linewidth]{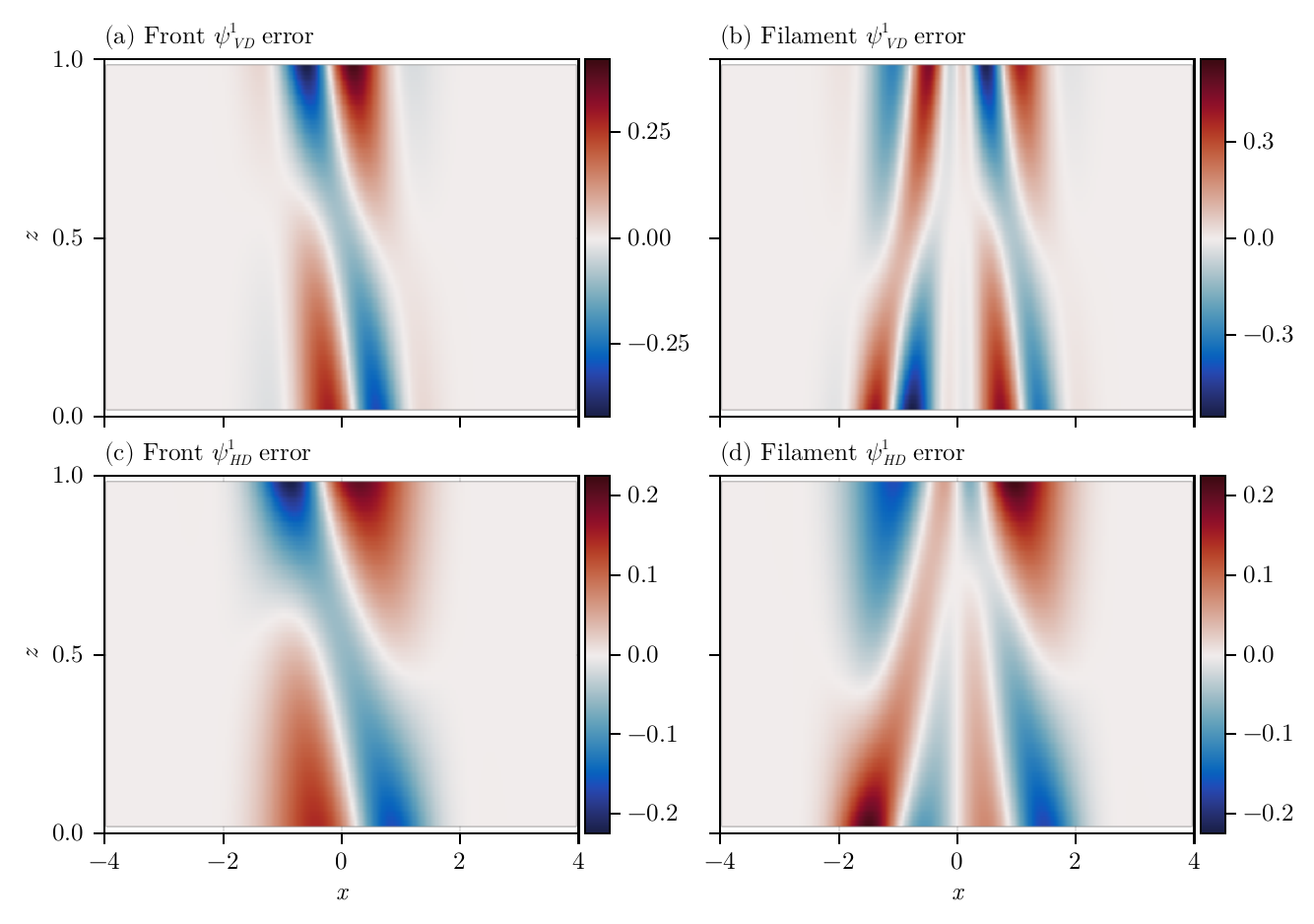}}
  \caption{Relative error of first-order streamfunctions due to boundary layer turbulence.}
\label{fig:psiHV_e}
\end{figure}

\section{Frontogenetic levels with viscosities and strain}\label{app:vis}
\begin{figure}[htbp]
  \centerline{\includegraphics[width=\linewidth]{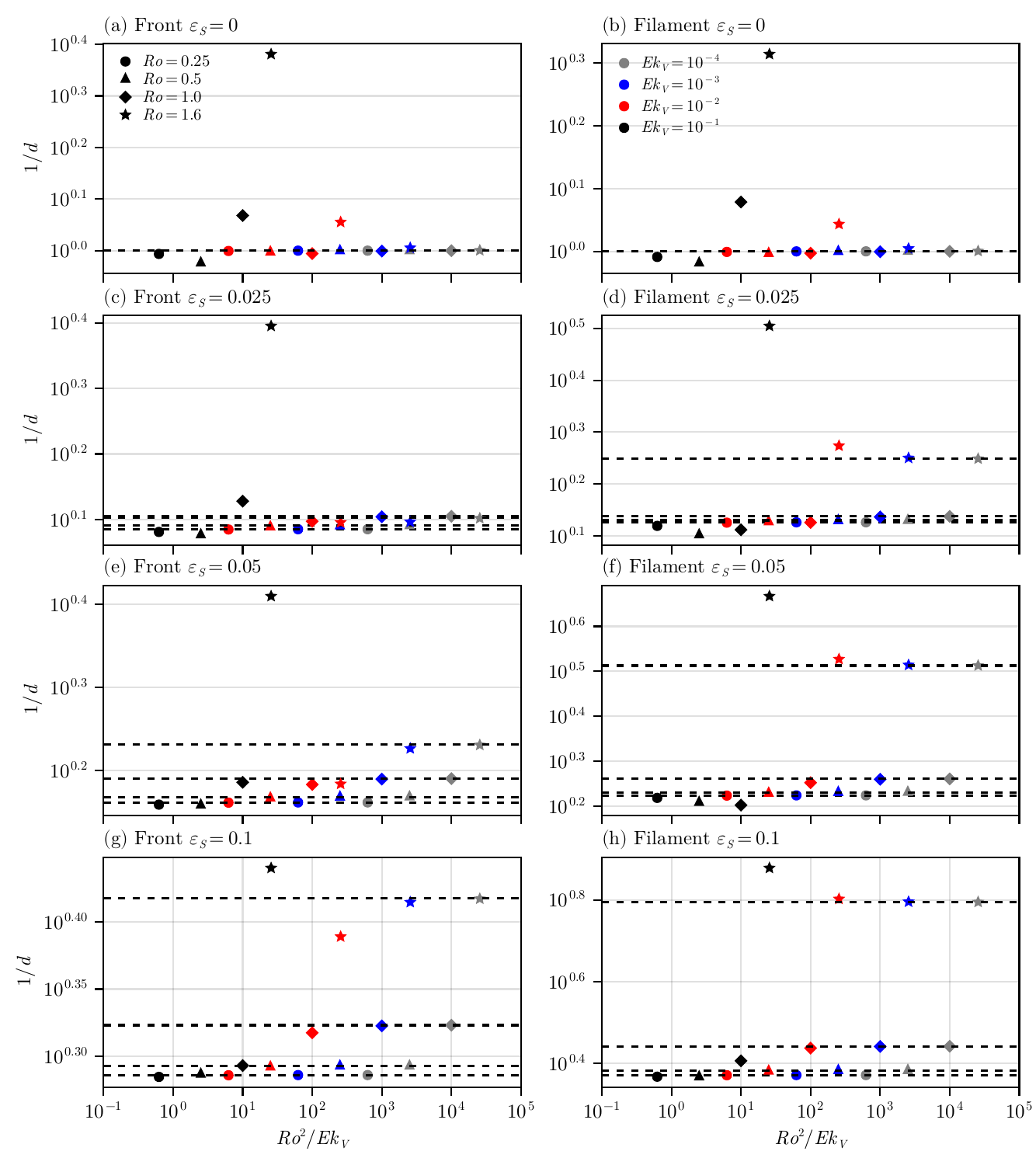}}
  \caption{Same as figure~\ref{fig:ro2ekV_sff} but with $\Pran_V=\infty$.}
\label{fig:ro2ekV_vsff}
\end{figure}

\begin{figure}[htbp]
  \centerline{\includegraphics[width=\linewidth]{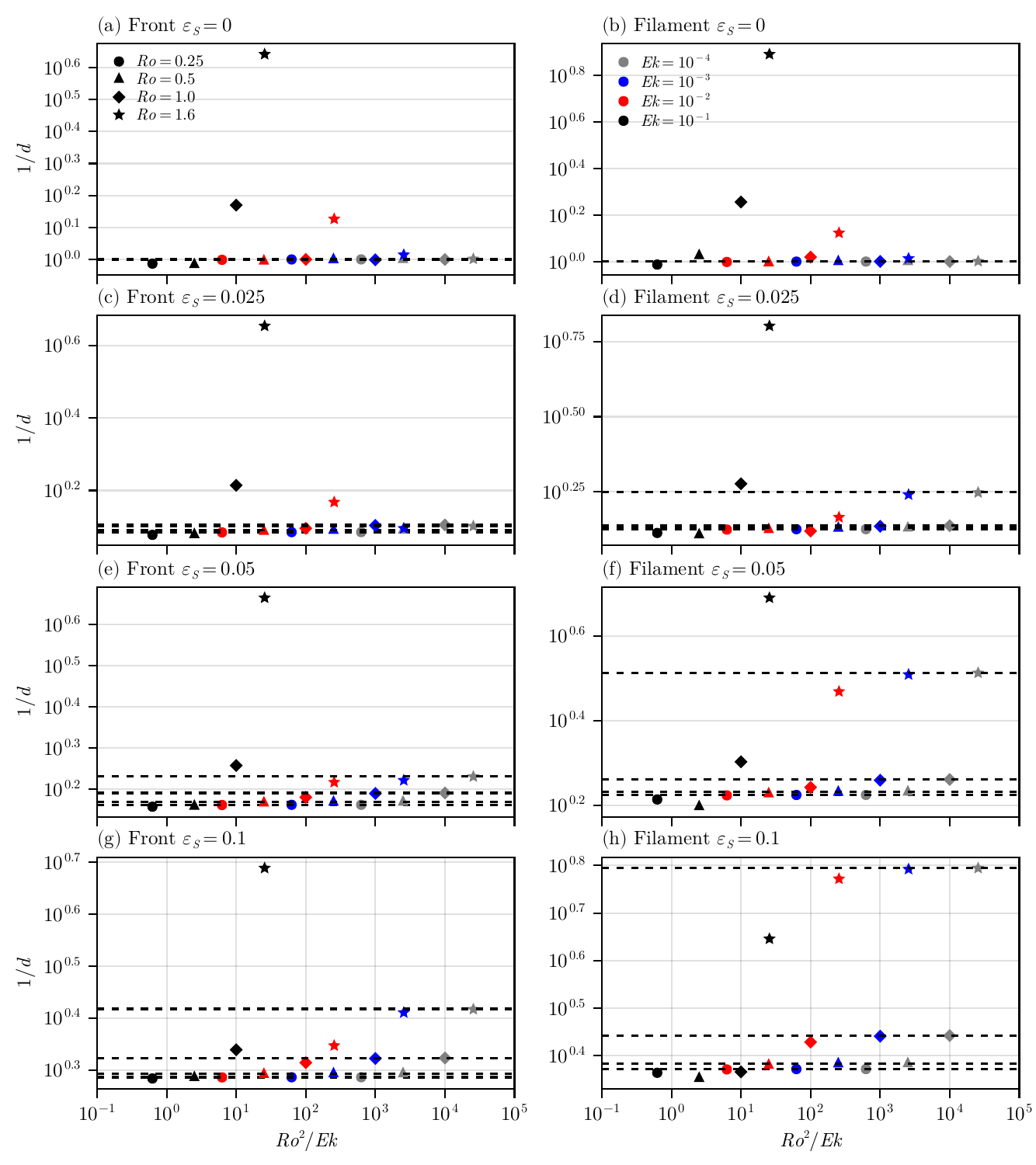}}
  \caption{Same as figure~\ref{fig:ro2ek_sff} but with $\Pran_V=\Pran_H=\infty$.}
\label{fig:ro2ek_vsff}
\end{figure}

\end{appen}\clearpage

\bibliographystyle{jfm}
\bibliography{jfm}

\end{document}